\documentclass[prb, aps, twocolumn, amsmath, amssymb, graphicx, longbibliography, nofootinbib]{revtex4-2}

\usepackage{bm}
\usepackage{graphicx}
\usepackage{braket}
\usepackage[normalem]{ulem}

\usepackage{color}
\usepackage[usenames, dvipsnames]{xcolor}
\usepackage[colorlinks=true]{hyperref}
\hypersetup{colorlinks=true, linkcolor=MidnightBlue, citecolor=MidnightBlue, urlcolor=OliveGreen}

\usepackage{apptools}
\usepackage[]{cleveref}
\Crefname{section}{Sec.}{Secs.}
\Crefname{subsection}{Sec.}{Secs.}
\Crefname{appendix}{\IfAppendix{Sec.}{Sec.}}{\IfAppendix{Secs.}{Secs.}}
\Crefname{subappendix}{\IfAppendix{Sec.}{Sec.}}{\IfAppendix{Secs.}{Secs.}}
\Crefname{equation}{Eq.}{Eqs.}
\Crefname{figure}{Fig.}{Figs.}
\Crefname{tabular}{Tab.}{Tabs.}

\newcommand{\figpanel}[1]{\textbf{#1}:}

\usepackage{siunitx}
\DeclareSIUnit{\ueV}{\micro\electronvolt}
\DeclareSIUnit{\um}{\micro\meter}
\DeclareSIUnit{\nm}{\nano\meter}
\usepackage{soul}
\usepackage{url}

\usepackage{tikz}

\usepackage{tabularx}
\usepackage{array}
\newcolumntype{I}{>{\hsize=0.36\hsize}X}
\newcolumntype{C}{>{\hsize=0.3\hsize\centering\arraybackslash}X}
\newcolumntype{U}{>{\hsize=0.35\hsize\centering\arraybackslash}X}
\newcolumntype{V}{>{\hsize=0.4\hsize\centering\arraybackslash}X}
\newcolumntype{W}{>{\hsize=0.44\hsize\centering\arraybackslash}X}
\newcolumntype{Y}{>{\hsize=0.49\hsize\centering\arraybackslash}X}

\usepackage{multirow}
\usepackage{makecell}
\usepackage[dvipsnames]{xcolor}

\newcommand{\Q}{\mathrm{\scriptscriptstyle Q}}
\newcommand{\Cq}{C_\Q}
\newcommand{\tCq}{{\tilde C}_\mathrm{\scriptscriptstyle Q}}

\newcommand{\ED}{E_\mathrm{\scriptscriptstyle D}}
\newcommand{\EM}{E_\mathrm{\scriptscriptstyle M}}
\newcommand{\EC}{E_\mathrm{\scriptscriptstyle C}}
\newcommand{\EJ}{E_\mathrm{\scriptscriptstyle J}}
\newcommand{\ECi}[1]{E_{{\rm\scriptscriptstyle C} #1}}
\newcommand{\HF}{H_\mathrm{\scriptscriptstyle F}}
\newcommand{\tHF}{{\tilde H}_\mathrm{\scriptscriptstyle F}}
\newcommand{\tauRTS}{\tau_\mathrm{\scriptscriptstyle RTS}}
\newcommand{\Svv}{S_{\scriptscriptstyle VV}}
\newcommand{\GammaEO}{\Gamma_{\scriptscriptstyle EO}}
\newcommand{\CPB}{\mathrm{\scriptscriptstyle CPB}}

\newcommand{\kB}{k_\mathrm{\scriptscriptstyle B}}

\renewcommand{\L}{\mathrm{\scriptscriptstyle L}}
\newcommand{\R}{\mathrm{\scriptscriptstyle R}}
\newcommand{\C}{\mathrm{\scriptscriptstyle C}}

\newcommand{\Bpara}{B_\parallel}
\newcommand{\Bperp}{B_\perp}

\newcommand{\VWP}{V_\mathrm{\scriptscriptstyle WP1}}
\newcommand{\QD}[1]{{{\mathrm{\scriptscriptstyle QD} #1}}}
\newcommand{\VQD}[1]{V_\QD{#1}}

\newcommand{\DeltaT}{\Delta_\mathrm{\scriptscriptstyle T}}

\DeclareMathOperator{\tr}{tr}
\DeclareMathOperator{\erf}{erf}
\DeclareMathOperator{\SNR}{SNR}

\newcommand{\DeltaInd}{\Delta_\mathrm{ind}}

\renewcommand{\ket}[1]{|#1 \rangle}

\newcommand{\tm}[1]{t_{\mathrm{m} #1}}
\newcommand{\Ng}[1]{N_{\mathrm{g} #1}}
\newcommand{\barNg}[1]{\bar{N}_{\mathrm{g} #1}}
\newcommand{\tC}{t_\C}
\newcommand{\tL}{t_\L}
\newcommand{\tR}{t_\R}

\renewcommand{\thesection}{\arabic{section}}
\renewcommand\thesubsection{\thesection.\arabic{subsection}}

\makeatletter
\def\Dated@name{Datum: }
\makeatother

\begin{document}

\title{Interferometric Single-Shot Parity Measurement in InAs-Al Hybrid Devices}

\author{Microsoft Azure Quantum$^\dagger$}


\begin{abstract}
The fusion of non-Abelian anyons or topological defects is a fundamental operation in measurement-only topological quantum computation. 
In topological superconductors, this operation amounts to a determination of the shared fermion parity of Majorana zero modes. 
As a step towards this, we implement a single-shot interferometric measurement of fermion parity in indium arsenide-aluminum heterostructures with a gate-defined nanowire.
The interferometer is formed by tunnel-coupling the proximitized nanowire to quantum dots.
The nanowire causes a state-dependent shift of these quantum dots' quantum capacitance of up to $\SI{1}{\femto\farad}$. 
Our quantum capacitance measurements show flux $h/2e$-periodic bimodality with a signal-to-noise ratio of $1$ in $\SI{3.7}{\micro\second}$ at optimal flux values. 
From the time traces of the quantum capacitance measurements, we extract a dwell time in the two associated states that is longer than $\SI{1}{\milli\second}$ at in-plane magnetic fields of approximately $\SI{2}{\tesla}$. 
These results are consistent with a measurement of the fermion parity encoded in a pair of Majorana zero modes that are separated by approximately $\SI{3}{\micro\meter}$ and subjected to a low rate of poisoning by non-equilibrium quasiparticles.
The large capacitance shift and long poisoning time enable a parity measurement error probability of $1\%$.
\end{abstract}

\maketitle


\section{Introduction}
\label{sec:introduction}

In order to leverage a topological phase for quantum computation, it is crucial to manipulate and measure the topological charge. 
This can be achieved through protected operations such as braiding and fusing non-Abelian anyons, which offer exponential suppression of errors induced by local noise sources and a native set of discrete operations~\cite{Kitaev97, Freedman98, Nayak08}.
Protocols for measurement-only topological quantum computation simplify these operations, reducing them to fusion alone~\cite{Bonderson08b, Bonderson08c, Karzig17}. This fundamental measurement is sufficient to enact all topologically protected operations. Novel error correction schemes have been developed
to take advantage of the operations available in measurement-only topological quantum computation
~\cite{Knapp18b, Hastings21, Paetznick23, Grans-Samuelsson23}. The robustness against errors and simplicity of control offered by this approach make measurement-based topological qubits a promising path towards utility-scale quantum computation, where managing the interactions of millions of qubits is necessary~\cite{Fowler12, Preskill18, Gidney21, Beverland22}.

One-dimensional topological superconductors (1DTSs)~\cite{Kitaev01, Lutchyn10, Oreg10} are a promising platform for building topological qubits~\cite{Lutchyn18}. 
Quantum information is stored in the fermion parity of Majorana zero modes (MZMs) localized at the ends of superconducting wires~\cite{Alicea12a, DasSarma15, Aasen16}, and projective measurements of the fermion parity are used to process quantum information and perform qubit state readout~\cite{Sau11a,Heck12,Hyart13}. 
The fermion parity shared by a pair of MZMs can be determined through an interferometric measurement~\cite{Fu09, Akhmerov09,Fu09a, Lutchyn10, Hassler10, Fu10, Heck11, Houzet13, Pientka13a, Cheng15, Sau15, Hell18, Chiu18, Drukier18, Liu19}.
A number of conceptual designs for topological qubits incorporate such interferometers~\cite{Fidkowski11b, Karzig17, Plugge16, Plugge17, Vijay15, Vijay16a, Vijay16b}.
Progress in this direction was made in Ref.~\onlinecite{Whiticar20}, which reported coherent transport through a gate-defined nanowire island in an Aharonov-Bohm interferometer, albeit in a regime that does not allow qubit readout.

\begin{figure}
\includegraphics[width=\columnwidth]{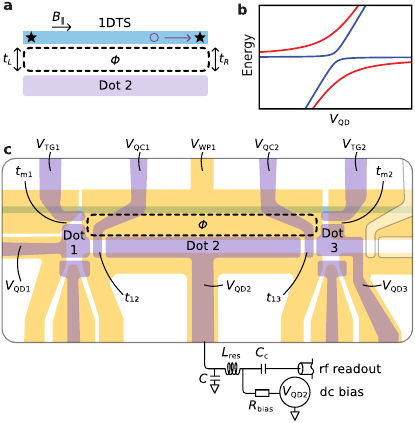}
\vspace{-5mm}
\caption{
\figpanel{a}~Minimal model of the system.
A nanowire tuned into a 1DTS state hosts MZMs at its ends, depicted by $\star$ symbols.
A quantum dot is tunably coupled to the MZMs by tunnel couplings $\tL$ and $\tR$ forming an interferometer (dashed lined) which is sensitive to the magnetic flux $\Phi$ and the combined fermion parity $Z$ of the dot-MZMs system. Poisoning by a quasiparticle (purple circle) flips the parity.
\figpanel{b}~Example energy spectra of the interferometer with total parity $Z=-1$ (red) and $Z=+1$ (blue) in the vicinity of the avoided crossing between the states with $N$ and $N+1$ electrons on the dot, as a function of the plunger voltage on the quantum dot. 
A measurement of the quantum capacitance, which probes the curvature of the energy spectrum at resonance, allows for rf readout of the parity.
\figpanel{c}~Gate layout for the interference loop formed by the triple-quantum-dot and the gate-defined nanowire. 
The effective couplings $t_L$ and $t_R$ of panel a depend on the couplings $\tm{1}$, $t_{12}$ and $\tm{2}$, $t_{23}$ and detuning of quantum dot~1 and 3, respectively.
Quantum dot~2 is capacitively coupled to an off-chip resonator chip for dispersive gate sensing and $\Cq$ measurement, which also includes a bias tee for applying dc voltages. 
See \Cref{fig:device_configuration} and \Cref{sec:scales} for a complete device schematic and gate naming convention; throughout the paper $V_i$ refers to the dc voltage applied to gate $i$.
}
\label{fig:minimal_model}
\end{figure}

In this paper, we demonstrate an interferometric measurement of the parity of a near-zero-energy state in a 1D nanowire, thereby validating a necessary ingredient of topological quantum computation. 
The measurement technique is based on probing the quantum capacitance $\Cq$ of a quantum dot coupled to the nanowire \cite{Karzig17, Plugge17, Munk20, Steiner20, Khindanov21a} and allows determination of the parity in a single shot (\Cref{fig:minimal_model}), with a probability of assignment errors of $1 \%$ for
optimal measurement time.
By itself, this measurement does not unequivocally distinguish between MZMs in the topological phase and fine-tuned low-energy Andreev bound states in the trivial phase~\cite{Janvier15, Hays18, Hays21, Wesdorp23, Elfeky23a}, but it does require the low-energy state to be supported at both ends of the wire and very weakly coupled to other low-energy states. 
Moreover, it provides a measurement of the state's energy with single-$\SI{}{\ueV}$ resolution. 
These features of the measurement strongly constrain the nature of the low-energy state. In a follow-up paper~\cite{mpr_tgp}, we will discuss correlations between $\Cq$ measurements and the topological gap protocol (TGP) phase diagram~\cite{Pikulin21, Aghaee23}.

\section{Topological qubit device design and setup}
\label{sec:device-design}

\begin{figure*}
\includegraphics[width=17.9cm]{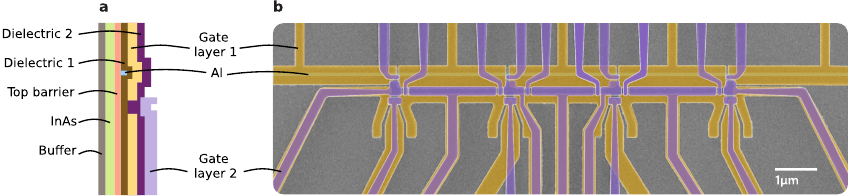}
\vspace{-4mm}
\caption{
\figpanel{a}~Cross-section of the topological qubit device design. 
\figpanel{b}~An SEM image with the aluminum strip (blue), first gate layer (yellow), and second gate layer (purple) indicated in false color.
}
\label{fig:device_design}
\end{figure*}

In this work, we introduce a topological qubit design that allows one to perform projective measurements of fermion parity encoded in MZMs. The device is composed of two primary components, as illustrated in \Cref{fig:minimal_model}. 
The first component is a nanowire, sections of which can be tuned into a 1DTS state, leading to topological degeneracy of the many-body ground state. 
The second component consists of quantum dots, which are designed to couple pairs of MZMs in an interferometric loop.
This device's 1DTS is based on a gated superconductor-semiconductor heterostructure~\cite{Suominen17, Nichele17, Poeschl22, Aghaee23}. 
The active region of the semiconductor consists of a $\SI{9.1}{\nano \meter}$ thick InAs quantum well with a $\SI{6}{\nano \meter}$ thick In$_{0.88}$Al$_{0.12}$As top barrier and $\SI{25}{\nano \meter}$ thick In$_{0.845}$Al$_{0.155}$As lower barrier, depicted schematically in \Cref{fig:minimal_model}c.
The superconductor is a $\SI{60}{\nano\meter}$ wide Al strip (blue in \Cref{fig:minimal_model,fig:device_design}) deposited with a thickness of $\SI{6.5}{\nano\meter}$ on the semiconductor. 
The material combination and dimensions have been optimized for values of the induced gap, spin-orbit coupling, and localization length that are favorable for the topological phase.
Device fabrication and the details of the heterostructure design are discussed in \Cref{sec:fabrication,sec:material-stack}, respectively.

In the full device, the nanowire is divided into 5 segments; one of them is shown schematically in \Cref{fig:minimal_model}c while all 5 are visible in \Cref{fig:device_design}b.
Each has a different ``plunger'' gate in the first gate layer (yellow in \Cref{fig:minimal_model,fig:device_design}) that controls the density in the corresponding region of the InAs quantum well.
To form a qubit, the second and fourth segments, each of length $L\approx\SI{3}{\micro\meter}$ long, need to be tuned into the topological phase while the other three need to be fully depleted underneath the Al nanowire (see \Cref{fig:device_configuration} in \Cref{sec:setup} for details). 
In this configuration, these trivial regions separate the two topological sections from each other and from Ohmic contacts at the ends of the Al nanowire~\cite{Menard19}. 
A full qubit device therefore consists of two 1DTS, each containing a pair of MZMs, separated by a trivial section in the middle, and three interferometers that are used to couple neighboring pairs of MZMs.
The complete gate layout, abbreviations for the different gates, and the voltage configuration for operation as a qubit are described in \Cref{sec:setup}.
Here, we focus on the left topological section of the device, shown in \Cref{fig:minimal_model}c, and implement a parity measurement using its associated interferometer. 

Our readout circuit is based on dispersive gate sensing of a triple quantum dot interferometer (TQDI): three electrostatically defined quantum dots that together with the 1DTS form a loop threaded by a flux, $\Phi$ (\Cref{fig:minimal_model}a,c). 
We control $\Phi$ by varying the out-of-plane magnetic field, $B_\perp$. 
The TQDI has two smaller dots (dots 1 and 3) which serve as tunable couplers and provide control over the tunnel couplings $\tL$ and $\tR$.  
The smaller dots are connected to the ends of the 1DTS through tunnel couplings $\tm{i}$, where $i=1,2$, and a long quantum dot (dot~2) that connects to dot~1 (dot~3) through tunnel couplings $t_{12}$ ($t_{23}$). 
The quantum capacitance, $\Cq$, of dot~2 is read out through dispersive gate sensing using an off-chip resonator circuit in a reflectometry setup~\cite{Colless13, Hornibrook14, Higginbotham14, Ahmed18, deJong19, vanVeen19, Sabonis19, Schaal20, Malinowski22a}. 
To improve the signal-to-noise ratio (SNR), a Josephson traveling wave parametric amplifier (JTWPA)~\cite{Macklin2015} is used in the first stage of amplification, followed by a low-noise high-electron-mobility transistor (HEMT) amplifier. 
Filtering is applied to dc lines as well as the rf readout lines to reduce the noise and radiation incident from higher temperature stages. Isolation on the rf lines is used to further reduce in-band back-action from the reflectometry setup. 
In addition to the line filtering and isolation, we employ multiple layers of shielding of the sample to suppress the generation of non-equilibrium quasiparticles (QPs) by stray infrared radiation~\cite{Barends11, Diamond22}. 
A detailed description of the reflectometry setup is given in~\Cref{sec:readout_system}.

Our TQDI device design addresses two crucial challenges. 
First, the device size is subject to conflicting requirements. 
To suppress the Majorana splitting $\EM \sim \DeltaT \exp(-L/\xi)$~\cite{Meng09}, we require $L\gg \xi$ (where $\xi$ is the disordered coherence length and $\DeltaT$ is the topological gap). 
However, an increase in $L$ also suppresses (albeit algebraically) the level spacing and charging energy of the dot, which suppresses the interference signal, as we discuss in \Cref{sec:linear-response,sec:toy-model-sims}. 
Our triple dot design offers a solution to these issues. 
With a length of $\SI{2.4}{\micro\meter}$, dot~2 retains a charging energy of $\approx \SI{60}{\ueV}$ deep in the Coulomb blockade regime, which is renormalized down to $\approx \SI{45}{\ueV}$ in the interferometer's operating regime. 
The level spacing is $\approx \SI{20}{\ueV}$ (see also \Cref{sec:dot-tuning}). 
To minimize the effect of disorder within dot~2, which may hinder elastic co-tunneling, we operate the dot with an occupation of $\sim 1000$ electrons. 
Dot~2 has significant effective tunneling matrix elements to both ends of the wire via the small dots of effective length $\SIrange{400}{500}{\nano\meter}$. 
The effective dot-to-wire couplings, $\tL$ and $\tR$, can be fine-tuned by adjusting the microscopic parameters that determine them: $t_{12}$, $t_{23}$, $\tm{1}$, $\tm{2}$, $\Ng{1}$, $\Ng{3}$.
The tunnel couplings, $t_{ij}, \tm{i}$ are indicated in \Cref{fig:device_design}b.
The dimensionless gate offset charge $\Ng{i}$ is controlled by the gate voltage $\VQD{i}$ and is given (up to an offset) by $\Ng{i} = \alpha_i e \VQD{i} / 2\EC$ where $\alpha_i$ is the lever arm of the associated gate, and $\EC$ is the charging energy of the dot.  
The relationship between the microscopic parameters and an effective model is further described in \Cref{sec:triple-dot-effective-H}.

The second challenge that this design addresses is the need for a substantial lever arm for QD2, which is the gate that couples the dot to the readout resonator. 
The signal amplitude is determined by the quantum capacitance, $\Cq^{(n)} = -e^2 \alpha^2 \,{\partial^2 {\varepsilon_n}}/{\partial \ED^2}$, for a state $|n\rangle$ in TQDI configuration with energy $\varepsilon_n$ with $\ED$ being the detuning of dot~2 from the charge degeneracy point.
We have optimized the lever arm by using a dual-layer gate geometry which enables us to place the plunger directly over the dot. As detailed in \Cref{sec:design-details}, our measurements confirm that QD2 has a lever arm $\alpha \approx \SIrange{0.4}{0.5}{}$, and our simulations are consistent with this range of values. 
Combined with our high bandwidth dispersive gate sensing setup and low-noise parametric amplifier that operates close to the quantum limit, this lever arm ensures that we can detect $\Cq$ with sufficient SNR. 
Readout can be activated by tuning all dots into resonance with the MZMs and by balancing the effective tunnel couplings between dot~2 and the MZMs, $\tL$ and $\tR$, to values which are comparable to or larger than the temperature.

Our device permits both dc and rf measurements, enabling the development of an rf-based QD-MZM tuning protocol
that we use to balance the arms of the interferometer. 
The protocol uses a measurement of $\Cq$ in a configuration where one of the small dots is maximally detuned to effectively interrupt the loop.
These measured quantities are fit to simulations to extract the couplings $t_{12}$, $t_{23}$, $\tm{1}$, and $\tm{2}$ (see \Cref{sec:toy-model-sims}). 
This measurement protocol expands upon the technique proposed in Refs.~\onlinecite{Clarke17, Prada17} and demonstrated in Ref.~\onlinecite{Deng16} which was based on dc transport measurements of the coupling between a quantum dot and a zero-energy state in a nanowire.
However, our rf-based protocol offers a finer resolution for the extraction of the couplings (down to single-$\SI{}{\ueV}$ level) and, thus, enables tuning the effective dot-to-wire couplings $\tL$ and $\tR$. 
Once we have determined the appropriate voltages for QD1 and QD3, we proceed with interferometer measurements. We can move through the bulk phase diagram of the nanowire by varying the in-plane field $B_\parallel$ and the voltage $\VWP$, indicated in, respectively, \Cref{fig:minimal_model}a,c. 
\Cref{sec:device-tune-up} contains further details of the tuneup procedure.
Complete details of tuning the nanowire into the topological phase will be discussed in Ref.~\onlinecite{mpr_tgp}.

\begin{figure*}
\includegraphics[width=17.9cm]{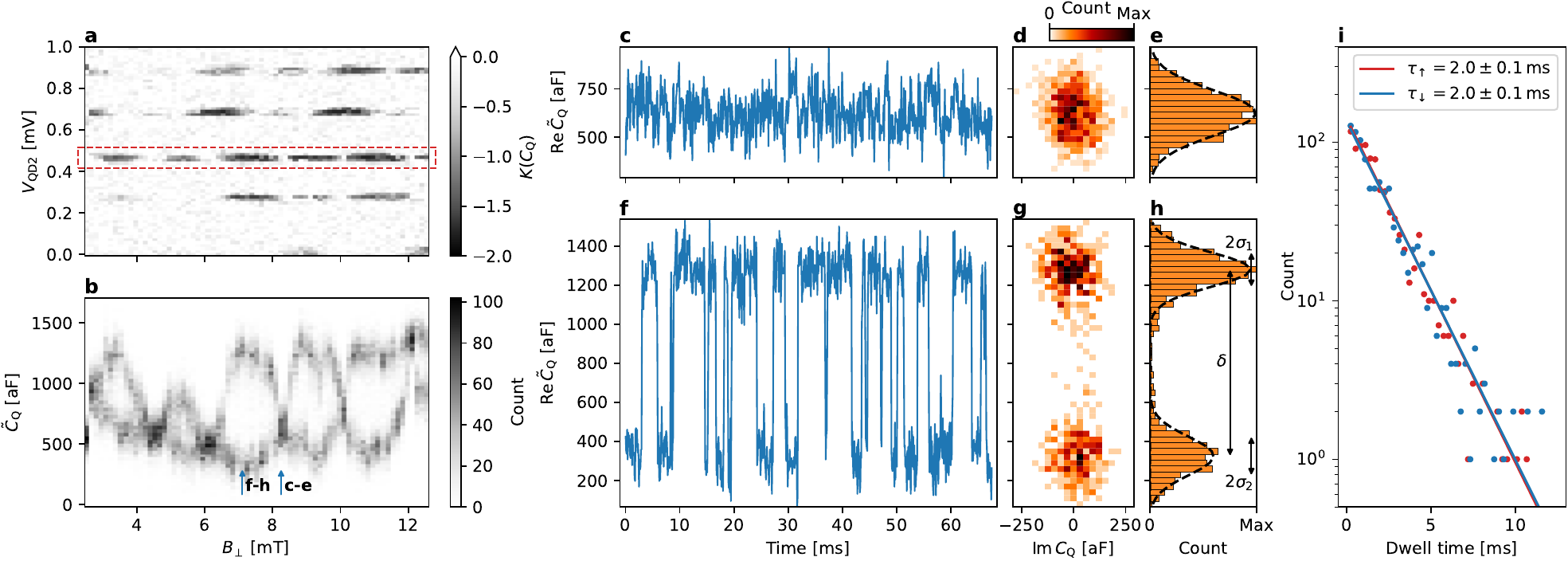}
\vspace{-5mm}
\caption{
\figpanel{a}~Kurtosis in the measured quantum capacitance, $K(\Cq)$, of dot~2 in device A (measurement A1) as a function of $B_\perp$ (which controls $\Phi$) and $\VQD{2}$ (which controls the dot~2 detuning) in the ${B_\parallel},\VWP$ parameter regime identified through the tune-up procedure discussed in the main text and \Cref{sec:device-tune-up}. 
\figpanel{b}~A histogram of $\tCq$ values as a function of flux for the $\VQD{2}$ value in the middle of the dashed red rectangle in panel a, showing clear bimodality that is flux-dependent with period $h/2e$.  
\figpanel{c,f}~Time traces at the two flux values marked by the vertical arrows in panel b, corresponding to minimal (panel c) and maximal (panel f) $\Delta\Cq$.
\figpanel{d,g}~The raw rf signal converted to complex $\tCq$ by the method described in \Cref{sec:cq_conversion} for the time trace shown in panels c and f. 
\figpanel{e,h}~Histograms of $\mathrm{Re}\,\tCq$ with Gaussian fits for an extraction of the $\mathrm{SNR} = \delta / (\sigma_1 + \sigma_2) = 4.9$,
the details of which are given in \Cref{sec:estimated-snr}.
\figpanel{i}~A histogram of dwell times aggregated over all values of $B_\perp$ where the signal shows bimodality. 
Fitting to an exponential shows that the up and down dwell times are both $\SI{2.0\pm 0.1}{\milli\second}$.
}
\label{fig:deviceA1_parity_measurements}
\end{figure*}

\section{Fermion parity measurement and interpretation}
\label{sec:parity_measurement}

To measure a time record of the fermion parity, we tune up the TQDI and perform a sequence of nearly $1.5 \times 10^4$ consecutive measurements of the resonator response, each with an integration time of $\SI{4.5}{\micro\second}$, thereby recording a time trace of total length $\SI{67}{\milli \s}$.
To improve visibility and compare with theoretical predictions we downsample the time trace to a resolution of $\SI{90}{\micro\second}$ and convert the resonator response into a $\Cq$ record using the procedure discussed in \Cref{sec:cq_conversion}.
We sweep the out-of-plane magnetic field $\Bperp$ in steps of $\SI{0.14}{\milli \tesla}$ to study the dependence on the external flux $\Phi$ through the interferometer loop and sweep $\VQD{2}$ to find charge transitions in dot~2. 
We use the kurtosis $K(\Cq)$ in the distribution of $\Cq$ values to detect bimodality. 
For a Gaussian distribution, the kurtosis satisfies $K=0$, while $K < 0$ indicates a bimodal distribution of two well-separated Gaussians; see \Cref{sec:kurtosis} for its definition.
One minor subtlety is that we plot $\tCq$, which includes an additive $\Bperp$-dependent contribution to the resonator response which is harmless because it cancels out of the quantites of interest $K(\Cq)$ and $\Delta\Cq$ (see the discussion below \Cref{eq:tilde-CQ-def} in \Cref{sec:cq_conversion} for details).

We tune dot~2 to charge degeneracy and use the TGP to select a magnetic field and a $\VWP$ range for our measurements. 
For device A, the relevant regime is in the neighborhood of $B_\parallel=\SI{1.8}{\tesla}$ and $\VWP = \SI{-1.832}{\volt}$.
When device A is in this regime, we observe oscillations in the kurtosis $K(\Cq)$ of the $\Cq$ time trace with a period of $\SI{1.7\pm0.2}{\milli \tesla}$ (shown for measurement A1 in \Cref{fig:deviceA1_parity_measurements}a). 
This period coincides with the $\SI{1.7}{\milli \tesla}$ expected to produce a flux of $h/2e$ through
the interference loop in this device geometry. 
The visibility and phase of the oscillations vary between successive charge transitions in dot~2.  
This observed behavior is consistent with the random matrix theory prediction for a disordered quantum dot~\cite{Aleiner02}. 
Indeed, a similar difference in the visibility of flux-induced oscillations across different charge transitions was recently observed in a double quantum dot interferometer experiment~\cite{Prosko23}.
In \Cref{sec:interferometry_regimes}, we discuss oscillations with different periods that are observed at other points in the device's parameter space.

Histograms of the $\tCq$ time trace for one of the dot~2 transitions in device A confirm that the negative kurtosis $K(\Cq)$ originates from a bimodal distribution of $\tCq$ values, as shown in \Cref{fig:deviceA1_parity_measurements}b. 
The time evolution of $\tCq$ exhibits a random telegraph signal (RTS) at flux values where there is negative kurtosis, as in \Cref{fig:deviceA1_parity_measurements}f, but no telegraph signal when the kurtosis is near zero, as in \Cref{fig:deviceA1_parity_measurements}c. 
As demonstrated in \Cref{fig:deviceA1_parity_measurements}i, the intervals between switches follow an exponential distribution with a characteristic time $\tauRTS \approx \SI{2}{\milli \s}$. 
From the histograms, we extract an achieved SNR of $4.9$ in $\SI{90}{\micro \second}$ (\Cref{fig:deviceA1_parity_measurements}g,h) or, equivalently, an SNR of $1$ in $\SI{3.7}{\micro \second}$ (see \Cref{sec:estimated-snr}). 
We interpret these $h/2e$-periodic bimodal oscillations and RTS in $\tCq$ as originating from switches of fermion parity. 
Such switches have been observed in mesoscopic superconducting devices, where they were triggered by non-equilibrium QPs infiltrating from the superconducting leads~\cite{Aumentado04, Naaman06, Lutchyn06, Ferguson06, Shaw08, Persson10, Barends11, Riste13, Janvier15, Hays18, Serniak19a, Uilhoorn21, Mannila22, Erlandsson23, Wesdorp23}.

We support this interpretation by reproducing our results with quantum dynamics simulations that incorporate rf drive power, charge noise, and temperature. 
To build intuition for those simulations, we use an idealized model (see \Cref{sec:triple-dot-effective-H}) subject to the follow assumptions: the wire is in the topological phase and there are no sub-gap states other than the MZMs; the charging energy and level spacing in the dots are much greater than the temperature; dot~1 and dot~3 are sufficiently detuned that their influence is fully encapsulated in the effective couplings $\tL$ and $\tR$ to MZMs at the ends of the wire (see \Cref{fig:minimal_model}a); and the drive frequency and power are both negligible. 
In this limit, the quantum capacitance as a function of the total fermion parity in the QD-wire system, $Z$, is given by 
\begin{multline}
\Cq(Z,\phi) 
= \frac{2 e^2 \alpha^2 |\tC(Z,\phi)|^2}
{\bigl[(\ED + 2Z \EM)^{2} + 4|\tC(Z,\phi)|^2\bigr]^{3/2}} \\
\times \tanh \biggl(\frac{\sqrt{(\ED + 2Z \EM)^{2} + 4|\tC(Z,\phi)|^2}}{2 \kB T}\biggr),
\label{eq:cq}
\end{multline}
where $\ED$ is the detuning from the charge degeneracy point, $\alpha$ is the lever arm of the plunger gate to the dot, $\EM$ is the MZM energy splitting, and $T$ is the temperature. 
The net effective tunneling that results from the interference between different trajectories from the dot to the MZMs and back, $\tC(Z,\phi)$, is
\begin{equation}
|\tC(Z,\phi)|^2 
= |\tL|^2 + |\tR|^2 + 2 Z |\tL| |\tR|  \sin\phi.
\label{eq:tc}
\end{equation}

Here, $\phi$ is the phase difference between $\tL$ and $\tR$, which is controlled by the magnetic flux $\Phi$ through the interference loop created by the dot, the wire, and the tunneling paths between them according to $\phi=2\pi\Phi/\Phi_0+\phi_0$, where $\Phi_0=h/e$ and $\phi_0$ is a flux-independent offset.
To capture the extent to which $\Cq$ can be used to discriminate between $Z=\pm 1$, it is convenient to introduce
\begin{equation}
\Delta\Cq(\phi) = |\Cq(Z=1,\phi) - \Cq(Z=-1,\phi)|.
\label{eq:delta_cq}
\end{equation}
The interferometer must be well-balanced $\tL \sim \tR$
in order for $\Delta\Cq$ to be large.
When $\EM = 0$, $\Delta\Cq$ exhibits maxima along the $\ED = 0$ line, with flux periodicity $h/2e$. In the presence of finite splitting $\EM \neq 0$, the $Z=1$ maxima form an $h/e$-periodic  arrangement along the $\ED = - 2 \EM$ line while the $Z=-1$ maxima form a similar arrangement along the $\ED = 2 \EM$ line, but out of phase by a flux offset of $h/2e$.

\begin{figure}
\includegraphics[width=\columnwidth]{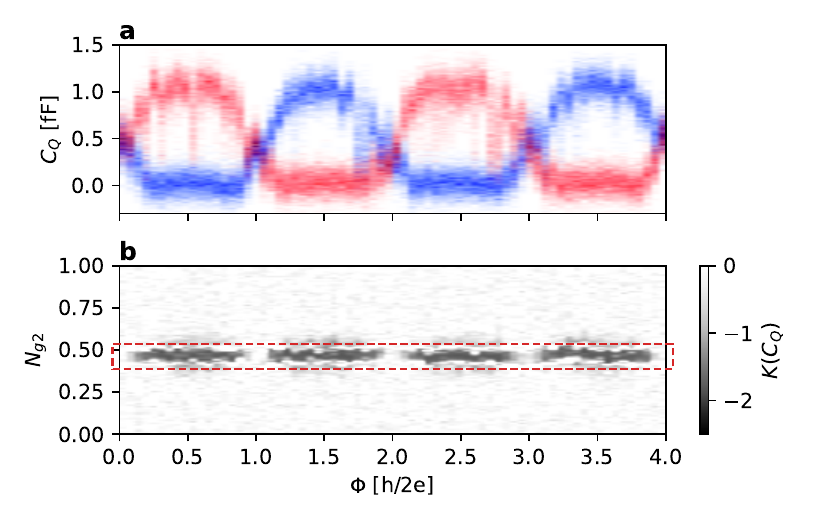}
\vspace{-8mm}
\caption{
Simulated dynamical $\Cq$ as a function of magnetic flux and dot~2 gate offset charge $\Ng{2}$, including the effects of charge and readout noise, as well as non-zero temperature, drive power, and frequency, per the discussion in the text. 
\figpanel{a}~Histogram of the two parity sectors for fixed $\Ng{2}=0.49$. Here, we used $\tm{1}=\tm{2}=\SI{6}{\ueV}$, $t_{12}=t_{23}=\SI{8}{\ueV}$, $\ECi{1}=\SI{140}{\ueV}$, $\ECi{2}=\SI{45}{\ueV}$, $\ECi{3}=\SI{100}{\ueV}$, $\Ng{1}=\Ng{3}=0.3$, $T=\SI{50}{\milli\kelvin}$, and $\EM = 0$.
\figpanel{b}~Kurtosis of $\Cq(t)$ as a function of $\Ng{2}$ and flux through the loop. The middle of the dashed red rectangle indicates the $\Ng{2}$ value used for the linecut in panel a.
}
\label{fig:simulated_response}
\end{figure}

For detailed comparison with experiments,  we simulate a more complete model of our interferometer, expanded to include the full triple-dot system, incoherent coupling to the environment, and backaction from the measurement. 
As before, we neglect all states in the wire except the MZMs.
Using the methods discussed in \Cref{sec:dynamical-cq,sec:toy-model-sims}, we compute $\Cq$ at temperature $T = \SI{50}{\milli\kelvin}$; charge-noise-induced dephasing rate $\gamma = \SI{1}{GHz}$; drive frequency $\omega=2\pi\times\SI{500}{MHz}$; detuning drive amplitude $A_\mathrm{rf} = \SI{5}{\ueV}$; and lever arm $\alpha = 0.45$. 
These values for the temperature and charge noise are based on quantum dot measurements in our system, which we discuss in~\Cref{sec:charge_noise,sec:electron_temperature}.

The simulated dynamical $\Cq$, defined in \Cref{sec:linear-response},
is shown in \Cref{fig:simulated_response}. The $\Cq$ histograms in \Cref{fig:simulated_response}a reveal two $h/e$-periodic branches (one shown in red and the other in blue), associated with the two parities of the coupled system.
For $\tm{1}=\tm{2}=\SI{6}{\ueV}$, $t_{12}=t_{23}=\SI{8}{\ueV}$, $\ECi{1}=\SI{140}{\ueV}$, $\ECi{2}=\SI{45}{\ueV}$, $\ECi{3}=\SI{100}{\ueV}$, $\Ng{1}=\Ng{3}=0.3$,
our simulations yield an estimated maximum
value as a function of flux
$\Delta\Cq \approx \SI{1}{\femto \farad}$, which is close to the measured value shown in \Cref{fig:deviceA1_parity_measurements}f-h.

If the fermion parity $Z$ were perfectly conserved, then the device would remain in one of the two parity eigenstates and the $\Phi$ dependence would follow either the blue or the red trace in \Cref{fig:simulated_response}a. 
However, $Z$ fluctuates on a time scale given by the quasiparticle poisoning time $\tau_\mathrm{qpp}$.  
Hence, in traces over times longer than $\tau_\mathrm{qpp}$, a bimodal distribution of $\Cq$ values is expected, i.e. \emph{both} the blue and red traces in \Cref{fig:simulated_response}a.
Consequently, the kurtosis $K(\Cq)$ exhibits minima where $\Delta\Cq$ is peaked, as shown in \Cref{fig:simulated_response}b, and time traces taken at these points will exhibit a telegraph signal composed of switches between the values $\Cq(1,\phi)$ and $\Cq(-1,\phi)$.

We can check the consistency of the time scale $\tau_\mathrm{qpp}$ by estimating the non-equilibrium quasiparticle density $n_\mathrm{qp}$ from the rate at which the combined QD-MZM system absorbs a non-equilibrium quasiparticle, setting $\tau_\mathrm{qpp} \equiv \tauRTS$, and verifying consistency with an independent, more direct, measurement. 
The rate of such absorption of a non-equilibrium quasiparticle can be estimated via $\tau_\mathrm{qpp}^{-1} = \gamma_0 \zeta n_\mathrm{qp} \mathcal{V}$, where $1/\gamma_0\sim \SI{1}{\nano\second}$ is a typical time scale for a quasiparticle to relax via electron-phonon coupling~\cite{Knapp18a} and become trapped in the topological wire segment, $\zeta$ is the probability  that an above-parent-gap quasiparticle is in the InAs forming the topological segment, $\mathcal{V}$ is the volume of the superconductor in contact with the topological segment, and $n_\mathrm{qp}$ is the density of quasiparticles in Al at $B_\parallel \sim \SI{2}{\tesla}$. 
The weight $\zeta$ is suppressed by the ratio of the semiconductor to superconductor densities of states and is typically of the order of $\zeta \sim 10^{-3}$ in the single subband regime~\cite{Aghaee23,Karzig21}. 
Within this model, poisoning time scales of $\tau_\mathrm{qpp}\sim \SI{2}{\milli\second}$ arise from a non-equilibrium quasiparticle density $n_\mathrm{qp}\sim \SI{1}{\micro\meter}^{-3}$.

In a separate experiment detailed in \Cref{sec:QPP}, we study quasiparticle poisoning using a Cooper pair box device which is filtered and shielded from radiation in a manner similar to the device in \Cref{fig:device_design} (see \Cref{{sec:readout_system}}). 
Using dispersive gate sensing, we measure the quantum capacitance of a Cooper pair box~\cite{Shaw08, Persson10} and observe even-odd switching events in real time. 
This enables us to extract the density of non-equilibrium quasi-particles in the Al strip, yielding $n_\mathrm{qp}^\CPB \approx \SI{0.6}{\micro\meter}^{-3}$ at $B_\parallel = 0$. 
This estimate is within an order of magnitude of the inferred value from $\tauRTS$ and is comparable to the densities measured in prior studies~\cite{Aumentado04, Naaman06, Lutchyn06, Ferguson06, Shaw08, Martinis09, Barends11, Riste13, Janvier15, Hays18, Serniak19a, Uilhoorn21, Mannila22, Erlandsson23, Wesdorp23}.

\begin{figure}
\includegraphics[width=6.5cm]{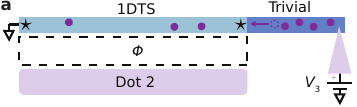}
\includegraphics[width=8.5cm]{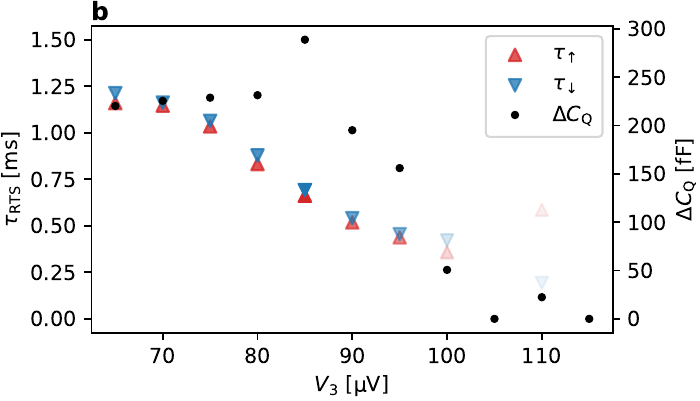}
\vspace{-2mm}
\caption{
\figpanel{a}~Schematic of the device configuration for the quasiparticle injection measurement.  The left interferometer loop is in the standard measurement configuration but now there is a trivial superconducting segment and a tunnel junction on the right half of the device to enable injection of current at an energy set by bias $V_3$.
\figpanel{b}~Extracted dwell times and $\Delta\Cq$ as a function of bias $V_3$.
For biases larger than $\approx \SI{70}{\micro\volt}$, we observe a decrease in dwell time $\tauRTS$, while the observed $\Delta\Cq$ remains stable up to about $\SI{90}{\micro\volt}$.
}
\label{fig:deviceB_dwell_times}
\end{figure}

The reproducibility of the observed phenomena is demonstrated in \Cref{sec:measurement-A2-B}, where we discuss a second measurement (A2) on device A (\Cref{fig:deviceA2_parity_measurements}) and a measurement of device B (\Cref{fig:deviceB1_parity_measurements}). 
We see very similar flux-dependent bimodality and RTS in these two additional data sets. 

We also perform three checks of our interpretation of the observed phenomena. First, we investigate the impact of injecting quasiparticles into the superconductor as indicated in \Cref{fig:deviceB_dwell_times}a (similar to the approach used in Ref.~\onlinecite{Elfeky23b}). 
Here, we make use of one of the tunnel junctions on the fourth gate-defined wire segment, which is not part of the interferometry loop. 
A fraction of the injected quasiparticles eventually reaches the MZMs, resulting in an enhanced switching rate.
As can be seen in \Cref{fig:deviceB_dwell_times}b (measured on device B), $\tauRTS$ decreases to half of its value as the bias voltage is increased from $0$ to $\SI{90}{\micro\volt}$,
while $\Delta \Cq$ exhibits weak dependence on bias voltage.
This indicates that quasiparticle injection primarily increases the rate of fermion parity switches, with minimal impact on other TQDI properties. 
For bias value of $\SI{100}{\micro\volt}$ or more, the switching time becomes comparable to the measurement time, making both $\Delta \Cq$ and $\tauRTS$ difficult to resolve.

Our second control experiment is to completely disconnect the dots from the wire. 
In devices A and B, we do not observe any RTS, as discussed in \Cref{sec:cutloop}.
This argues against the RTS being caused by two-level systems outside the wire (e.g. in the dielectric), an effect which has been observed in quantum dots and quantum point contacts~\cite{Cobden92, Peters99, Pioro05, Buizert08}.

Finally, we repeat the interferometry measurement at a low field ($\SI{0.8}{T}$), where the device is well within the trivial phase, and the induced gap in the nanowire has not closed yet. 
As shown in \Cref{sec:trivial_measurement}, we do not observe any $h/e$ flux periodicity as expected for a gapped wire, corroborating the expectation that the observed phenomena summarized in \Cref{fig:deviceA1_parity_measurements} is due to a low-energy state in the nanowire and that two-electron processes are negligible.

In \Cref{sec:quasi-MZM}, we extend the model introduced above to allow us to analyze the quasi-MZM scenario discussed in previous works~\cite{Prada12, Kells12, Tewari14, Liu17, Vuik19, Pan21b}.
We introduce an additional pair of ``hidden'' Majorana modes that are weakly coupled to each other and to the MZMs, which themselves are coupled to QD1 and QD3. 
While this scenario can manifest in the trivial phase, it is generic for the couplings to be large and some amount of fine-tuning is required to make them small.
We find that when these ``hidden'' Majorana modes are weakly coupled to each other and to other MZMs, the quantum capacitance signal is significantly suppressed, to a degree that is inconsistent with the measured values.
Our analysis suggests that a rather substantial coupling (larger than the temperature) between these ``hidden'' Majorana modes is necessary to account for the experimental results. 
In such cases, the ``hidden'' Majorana modes effectively become gapped out, bringing us back to the low-energy model described by \Cref{fig:minimal_model} [see \Cref{eq:HF}], which we compare to our measurements in this section [as in \Cref{eq:cq} and \Cref{fig:simulated_response}] and the next.

\section{Different observed interferometry regimes}
\label{sec:interferometry_regimes}

\begin{figure*}
\includegraphics[width=17.9cm]{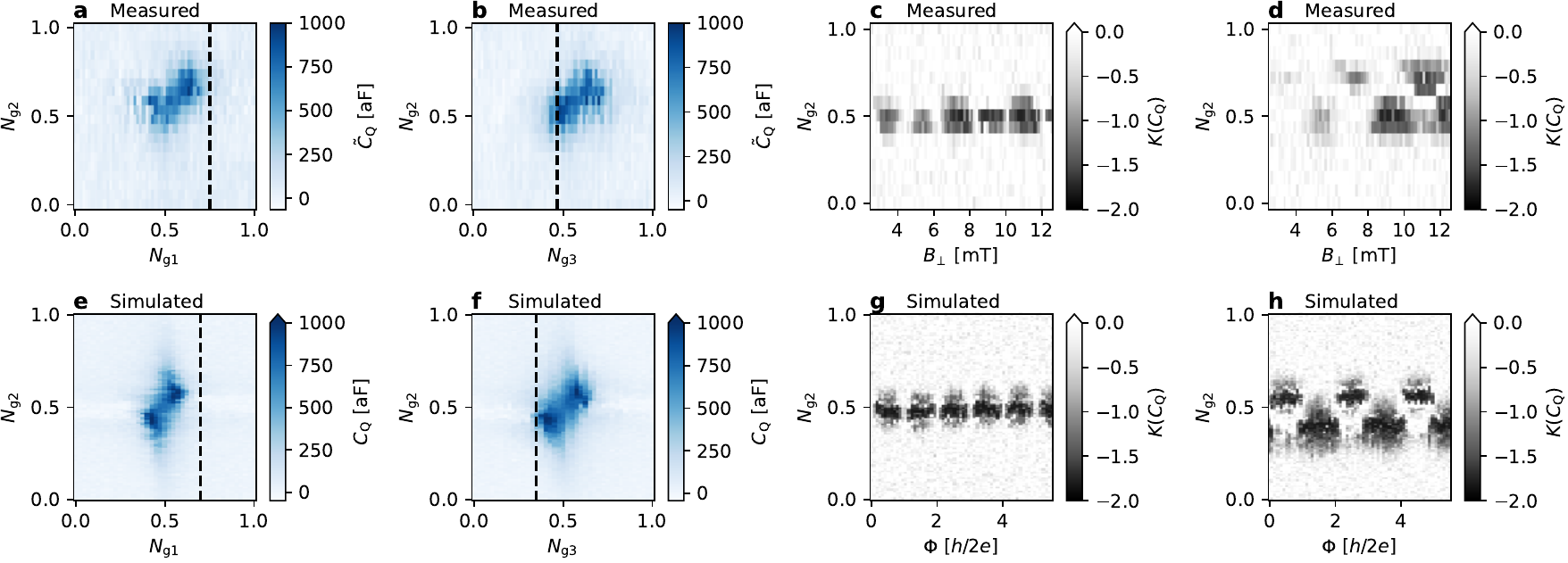}
\vspace{-6mm}
\caption{
\figpanel{a,b}~The measured gate-gate scans used to extract the QD-MZM couplings and (c) the measured kurtosis in the TQDI configuration, taken at the same point in the $(\Bpara, \VWP)$ phase diagram. The dashed lines indicate the points in $\Ng{1}$ and $\Ng{3}$ where the data in panels c and d were taken. 
In panels a and b, for easier comparison to the simulated data, we have shifted $\tCq$ to vanish deep in Coulomb blockade.
\figpanel{d}~The kurtosis $K(\Cq)$ with $\VWP$ increased by $\SI{0.39}{\milli\volt}$ relative to panel c.
Note that the bimodality is not visible at small $B_\perp$ here due to the movement of the resonator frequency with field.
\figpanel{e,f}~Simulated gate-gate scans  which are fit to the experimental data in panels a and b in order to extract the QD-MZM couplings.
\figpanel{g}~The simulated kurtosis for the same couplings and with $\EM = 0$.
\figpanel{h}~The simulated kurtosis in the topological phase with only $\EM$ modified to $\EM = \SI{3}{\ueV}$.
}
\label{fig:qd_mzm}
\end{figure*}

The oscillations shown in \Cref{fig:deviceA1_parity_measurements}b are the easiest to interpret but they are not the only type of oscillations that we observe when we vary the voltage $\VWP$ by $\SIrange{1}{2}{\milli\volt}$ around the configuration where we observe $h/2e$ oscillations. 
In this section, we show how these other cases can be explained within the model discussed in the previous section (see \Cref{sec:design-details} for more details of the model).

To apply this model, we first extract the model parameters $\tm{1}$, $\tm{2}$, $t_{12}$, and $t_{23}$ (along with the charges $\Ng{i}$ and charging energies). 
We fit the measured response consistently in three different configurations: (i)~dot~3 is maximally detuned, (ii)~dot~1 is maximally detuned, and (iii)~all dots are optimally tuned to balance the interferometer.
As an example, the scan shown in \Cref{fig:qd_mzm}a, which is taken with dot~3 detuned to cut off the right arm of the interferometer, can be quantitatively compared to the simulated data in \Cref{fig:qd_mzm}e to determine the couplings $t_{12}$ and $\tm{1}$. 
The extent of the dark blue region in the $\Ng{1}$ direction in \Cref{fig:qd_mzm}a is correlated with $\tm{1}$ while the extent along the $\Ng{1} - \Ng{2}$ diagonal is correlated with $t_{12}$ (see \Cref{sec:toy-model-sims} for details).
Similarly, \Cref{fig:qd_mzm}b and \Cref{fig:qd_mzm}f (taken with the left arm of the interferometer cut) are used to determine the couplings $t_{23}$ and $\tm{2}$. 
This leaves us with one independent parameter $\EM$, which can be fitted to $K(\Cq)$ data when both arms of the interferometer are connected.  

\begin{figure*}
{\hspace{-3mm}\includegraphics[width=18.2cm]{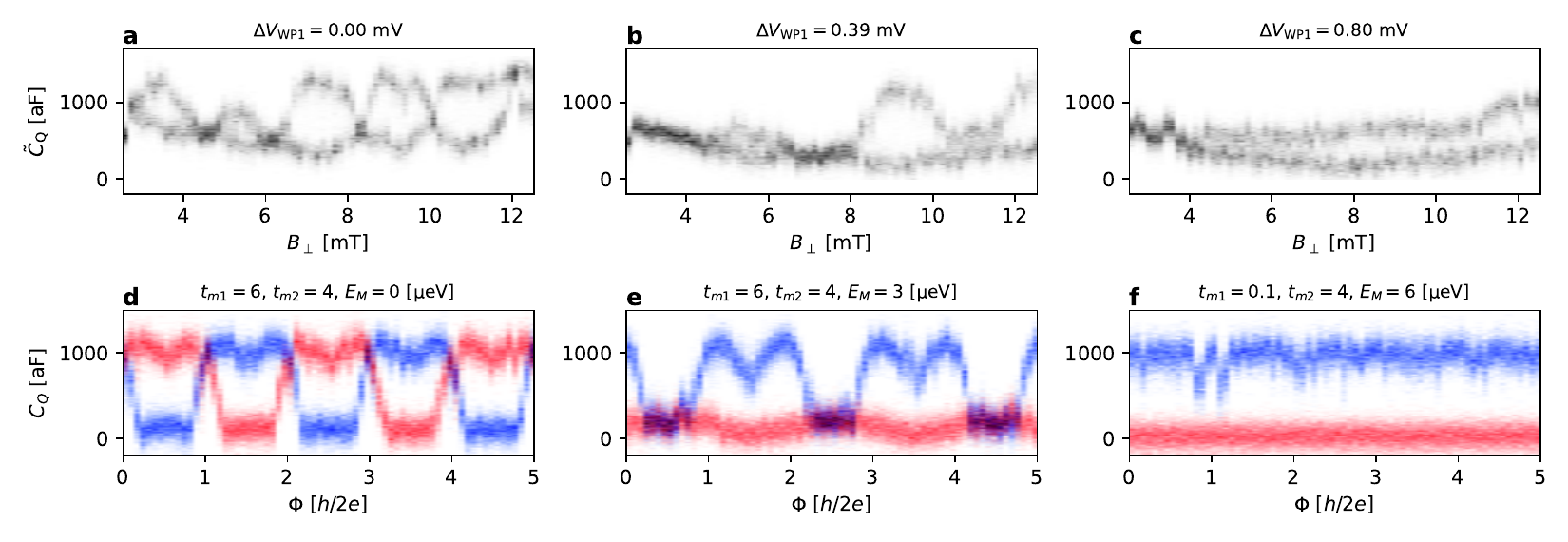}}
\vspace{-8.5mm}
\caption{
\figpanel{a-c}~Evolution of the histogram of resonator response values with voltage $V_\mathrm{WP1} = \SI{-1.8312}{\volt}+\Delta V_\mathrm{WP1}$, which is a sampling of a dataset that has more closely-spaced $\VWP$ values. 
We use the same greyscale as in \Cref{fig:deviceA1_parity_measurements}. 
Moving from left to right, panels a, b, and c exemplify scenarios a, b, and c described in the text. Each panel corresponds to a fixed charge transition of dot~2, typically the one with strongest signal.
\figpanel{d-f}~Simulated data illustrating how this evolution can be understood in terms of the splitting of the fermionic mode $\EM$ and the balancing of the couplings $\tm{1}$, $\tm{2}$. 
Here, as in \Cref{fig:qd_mzm}, blue and red indicate different total parities of the system. 
We have fixed $t_{12}=t_{23}=\SI{12}{\ueV}$, $\Ng{1}=0.7$ and $\Ng{3}=0.35$ throughout. 
Moving from left to right, the MZM splitting $\EM$ increases from 0 to $3$ to $\SI{6}{\ueV}$. 
Meanwhile the asymmetry between the left and right arms of the interferometer, $\tm{2}/\tm{1}$, is $1.5$ in panels d and e and $40$ in panel f.
}
\label{fig:scenarios}
\end{figure*}

We now classify the various types of oscillations that we have observed according to the following scenarios:
\begin{enumerate}
\item[(a)] The kurtosis $K(\Cq)$ is flux-dependent with a period of $h/2e$, as illustrated in \Cref{fig:qd_mzm}c. The histogram of $\tCq$ values shows bimodality with this period, as shown in \Cref{fig:deviceA1_parity_measurements}b and \Cref{fig:scenarios}a. We interpret this scenario as resulting from a balanced interferometer with two parity branches, both of which exhibit clear $h/e$ periodicity. 
\item[(b)] The kurtosis $K(\Cq)$ exhibits a characteristic $h/e$-periodic zig-zag pattern, as shown in \Cref{fig:qd_mzm}d. A fixed-$\VQD{2}$ horizontal cut through the $K(\Cq)$ plot has $h/e$-periodicity, and the histogram of $\tCq$ values shows bimodality with $h/e$-periodic oscillations that are more pronounced in one branch, as in \Cref{fig:scenarios}b. We identify this scenario with a balanced interferometer with two parity branches, one of which has stronger flux dependence than the other at a fixed value of $\Ng{2}$.  
\item[(c)] There is bimodality, but neither $K(\Cq)$ nor the histogram of $\tCq$ values shows visible periodicity in flux, as shown in \Cref{fig:scenarios}c. This scenario can be interpreted as a very unbalanced interferometer, resulting in an absence of distinct flux dependence in either of the parity branches. 
\end{enumerate}

We begin by explaining scenario (a) in terms of a balanced interferometer model with a small $\EM$ value. 
Applying the procedure described above to the dataset discussed in \Cref{fig:qd_mzm}a-c, we obtain the following model parameters for the particular charge transition point in that figure: $\tm{1} \approx \SI{6}{\ueV}$, $\tm{2} \approx \SI{4}{\ueV}$, $t_{12} \approx t_{23} \approx \SI{12}{\ueV}$, and $\EM \approx 0$. 
These parameters lead us to \Cref{fig:qd_mzm}e-g and \Cref{fig:scenarios}d. A close examination reveals good agreement between the measured (\Cref{fig:scenarios}a) and simulated (\Cref{fig:scenarios}d) data. 
Hence, scenario (a) can be explained by $\tm{2}\sim\tm{1}$ and small $\EM$.

In order to delve into scenario (b), we analyze the dataset displayed in \Cref{fig:scenarios}b (corresponding to a cut at $\Ng{2}=0.5$ in \Cref{fig:qd_mzm}d) and \Cref{fig:scenarios}e. 
Oscillations with period $h/e$ are apparent in one of the parity branches in \Cref{fig:scenarios}b%
\footnote{Note that $K(\Cq)$ in \Cref{fig:qd_mzm}d and the $\Delta\Cq$ oscillations in \Cref{fig:scenarios}b are suppressed for small $B_\perp$ because the resonator frequency shifts with $B_\perp$, which leads to a reduction in SNR.}
and e, thereby realizing scenario (b). 
This dataset is offset by $\Delta\VWP = \SI{0.39}{\milli\volt}$ from the data in \Cref{fig:scenarios}a. 
Our simple model reproduces similar data, as shown in \Cref{fig:qd_mzm}h and \Cref{fig:scenarios}e by taking $\EM = \SI{3}{\ueV}$ and keeping the other junction and dot parameters the same. 
This results in the characteristic zig-zag pattern in $K(\Cq)$ with peaks offset from each other by $4\EM$. 
In this regime, the observation of oscillations in both parity branches is unlikely, as can be seen in \Cref{fig:scenarios}e. 
The qualitative agreement apparent between \Cref{fig:scenarios}b and e suggests that scenario (b) is consistent with $\tm{2}\sim\tm{1}$ and a moderate $\EM$. 
It is worth noting that this method enables us to probe the Majorana splitting energy $\EM$ with single-$\SI{}{\ueV}$ resolution. 
This is a crucial parameter that characterizes the topological phase. 
Achieving a similar resolution for extracting $\EM$ in transport measurements is quite challenging~\cite{DasSarma12, Albrecht16, Vaitieknas20}.

Lastly, scenario (c) is depicted in \Cref{fig:scenarios}c. 
Here, $\tCq$ exhibits bimodality but lacks flux-dependent oscillations. 
This observation is consistent with $\tm{2}\gg\tm{1}$ and large $\EM$, as demonstrated in \Cref{fig:scenarios}f. 
In this configuration, the interferometer loop is almost severed, leading to the absence of flux dependence in both parity branches. 
Despite this, the bimodality is still present due to non-zero $\EM$, as may be seen in \Cref{fig:scenarios}f.
If the loop were severed and $\EM=0$, then the bimodality would disappear as well.

The changes we observe in the measured signal as a function of $\VWP$ can be attributed to changes in the properties of the low-energy state in the wire. We have fit our data to a model in which this state is assumed to be due to Majorana zero modes and find good agreement.
Within the topological phase, we expect a substantial variation in $\EM$ in our current devices, attributable to mesoscopic fluctuations due to finite-size effects. Indeed, in the current generation of devices~\cite{Aghaee23}, we anticipate modest energy gaps $\DeltaT$ and coherence lengths that are not significantly smaller than the device length. However, for devices with significantly higher $\DeltaT$, 
such variation in $\EM$ is expected only at the boundaries of the topological phase where the system turns gapless. Conversely, deep within the topological phase, robust $h/2e$-periodic oscillations of $K(\Cq)$,
indicative of bimodality, should be observable.

\section{Discussion and outlook}
\label{sec:discussion}

We have presented dispersive gate sensing measurements of the quantum capacitance in topological qubit devices designed for the readout of fermion parity shared between MZMs at the opposite ends of a nanowire. 
After tuning the nanowire density and in-plane magnetic field into the parameter regime where we expect the topological phase \cite{Aghaee23}, and balancing the interferometer formed by the nanowire and the quantum dots, we observed a flux-dependent bimodal random telegraph signal (RTS) in the quantum capacitance.
We interpret this RTS as switches of the parity of a fermionic state in the wire.
The long switching time $\tauRTS > \SI{1}{\milli\second}$ suggests a low quasiparticle poisoning rate,
which we find to be within an order of magnitude of the quasiparticle density measured in a Cooper pair box device. This interpretation has been further
validated by the
decrease of $\tauRTS$ that occurs
when we intentionally inject quasiparticles into the device and also by the disappearance of the RTS when we isolate the wire from the dots.
We have fit these data to a model in which the fermion parity is associated with two MZMs localized at the opposite ends of a 1DTS, and we find good agreement.
These measurements do not, by themselves, determine whether the
low-energy states detected by interferometry are topological. However, by fitting to a model of trivial Andreev
states, we have tightly constrained the properties that such
states would have to have in order to be consistent with
our data. To fully resolve this issue,
we will discuss the device's phase diagram and the stability of the observed flux-dependent RTS in a separate publication~\cite{mpr_tgp}.

In conclusion, our findings represent significant progress towards the realization of a topological qubit based on measurement-only operations. Single-shot fermion parity measurements are a key requirement for a Majorana-based topological quantum computation architecture.

\acknowledgements
We thank Haim Beidenkopf, Sankar Das Sarma, Leonid Glazman, Bertrand Halperin, Angela Kou, Katherine Moler, Wolfgang Pfaff, and Mark Rudner for discussions.
We thank Edward Lee and Todd Ingalls for assistance with
the figures. We are grateful for the contributions
of Anand Dokania, Alexandra Efimovskaya, Linda Johansson, and Andrew Mullally at an early stage of
this project. We have benefited from interactions with
Paul Accisano, Parsa Bonderson, Jan Borovsky, Tom Brown, Gary Campbell, Srivatsa Chakravarthi, Kushal Das, Neil Dick, Raghu Gatta, Haris Gavranovic, Michael Goulding, Jonathan Knoblauch, Sarah Jablonski, Seth Kimes, Jamie Kuesel, Jason Lee, Jake Mattinson, Ali Moini, Tim Noonan, Diego Olivier Fernandez Pons, Len Sanderson, Marcus P. da Silva, Patrick Strøm-Hansen, Satoshi Suzuki, Matt Turner, Richard Yu, Andrew Zimmerman.

\vspace{1mm}
\textbf{Correspondence and requests for materials} should be addressed to Chetan Nayak~(cnayak@microsoft.com).

\vspace{1cm}

$^\dagger${\small
Morteza Aghaee, Alejandro Alcaraz Ramirez, Zulfi Alam, Rizwan Ali, Mariusz Andrzejczuk, Andrey Antipov, Mikhail Astafev, Amin Barzegar, Bela Bauer, Jonathan Becker, Umesh Kumar Bhaskar, Alex Bocharov, Srini Boddapati, David Bohn, Jouri Bommer, Leo Bourdet, Arnaud Bousquet, Samuel Boutin, Lucas Casparis, Benjamin James Chapman, Sohail Chatoor, Anna Wulff Christensen, Cassandra Chua, Patrick Codd, William Cole, Paul Cooper, Fabiano Corsetti, Ajuan Cui, Paolo Dalpasso, Juan Pablo Dehollain, Gijs de Lange, Michiel de Moor, Andreas Ekefjärd, Tareq El Dandachi, Juan Carlos Estrada Salda\~na, Saeed Fallahi, Luca Galletti, Geoff Gardner, Deshan Govender, Flavio Griggio, Ruben Grigoryan, Sebastian Grijalva, Sergei Gronin, Jan Gukelberger, Marzie Hamdast, Firas Hamze, Esben Bork Hansen, Sebastian Heedt, Zahra Heidarnia, Jesús Herranz Zamorano, Samantha Ho, Laurens Holgaard, John Hornibrook, Jinnapat Indrapiromkul, Henrik Ingerslev, Lovro Ivancevic, Thomas Jensen, Jaspreet Jhoja, Jeffrey Jones, Konstantin V. Kalashnikov, Ray Kallaher, Rachpon Kalra, Farhad Karimi, Torsten Karzig, Evelyn King, Maren Elisabeth Kloster, Christina Knapp, Dariusz Kocon, Jonne Koski, Pasi Kostamo, Mahesh Kumar, Tom Laeven, Thorvald Larsen, Jason Lee, Kyunghoon Lee, Grant Leum, Kongyi Li, Tyler Lindemann, Matthew Looij, Julie Love, Marijn Lucas, Roman Lutchyn, Morten Hannibal Madsen, Nash Madulid, Albert Malmros, Michael Manfra, Devashish Mantri, Signe Brynold Markussen, Esteban Martinez, Marco Mattila, Robert McNeil, Antonio B. Mei, Ryan V. Mishmash, Gopakumar Mohandas, Christian Mollgaard, Trevor Morgan, George Moussa, Chetan Nayak, Jens Hedegaard Nielsen, Jens Munk Nielsen, William Hvidtfelt Padkær Nielsen, Bas Nijholt, Mike Nystrom, Eoin O'Farrell, Thomas Ohki, Keita Otani, Brian Paquelet Wütz, Sebastian Pauka, Karl Petersson, Luca Petit, Dima Pikulin, Guen Prawiroatmodjo, Frank Preiss, Eduardo Puchol Morejon, Mohana Rajpalke, Craig Ranta, Katrine Rasmussen, David Razmadze, Outi Reentila, David J. Reilly, Yuan Ren, Ken Reneris, Richard Rouse, Ivan Sadovskyy, Lauri Sainiemi, Irene Sanlorenzo, Emma Schmidgall, Cristina Sfiligoj, Mustafeez Bashir Shah, Kevin Simoes, Shilpi Singh, Sarat Sinha, Thomas Soerensen, Patrick Sohr, Tomas Stankevic, Lieuwe Stek, Eric Stuppard, Henri Suominen, Judith Suter, Sam Teicher, Nivetha Thiyagarajah, Raj Tholapi, Mason Thomas, Emily Toomey, Josh Tracy, Michelle Turley, Shivendra Upadhyay, Ivan Urban, Kevin Van Hoogdalem, David J. Van Woerkom, Dmitrii V. Viazmitinov, Dominik Vogel, John Watson, Alex Webster, Joseph Weston, Georg W. Winkler, Di Xu, Chung Kai Yang, Emrah Yucelen, Roland Zeisel, Guoji Zheng, Justin Zilke
}

\vspace{0.5 cm}




\bibliography{mpr}

\clearpage
\appendix
\newcounter{appendix}
\setcounter{appendix}{1}

\onecolumngrid
\section*{Supplementary information}
\twocolumngrid

\setcounter{section}{0}
\renewcommand{\thesection}{S\arabic{section}}
\renewcommand{\thesubsection}{S\arabic{section}.\arabic{subsection}}
\makeatletter
\renewcommand{\p@subsection}{}
\makeatother

\counterwithin{equation}{appendix}
\renewcommand{\theequation}{S\arabic{equation}}

\counterwithin{figure}{appendix}
\setcounter{figure}{0}
\renewcommand{\thefigure}{S\arabic{figure}}

\counterwithin{table}{appendix}
\renewcommand{\thetable}{S\arabic{table}}

\section{Device design, fabrication, and system setup}
\label{sec:setup}

\subsection{Device layout details}
\label{sec:scales}

\begin{figure*}
\includegraphics[width=17.5cm]{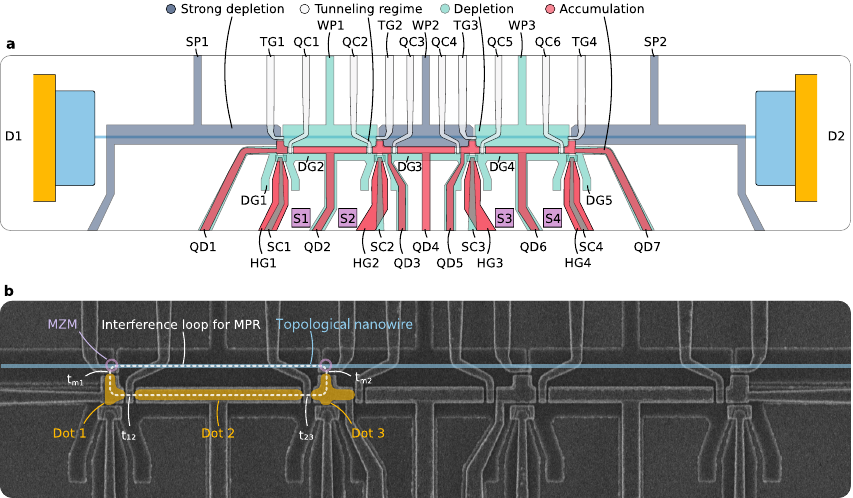}
\caption{
\figpanel{a}~The gate layout of our device and voltage ranges necessary to form the interferometry loops needed for a functioning qubit device.  
Normal metal drain contacts (D1 and D2) are indicated in yellow.  
Normal metal source contacts S1-S4 are formed in 2DEG regions indicated by the purple boxes.  
The dark gray gates are normally set to very negative voltages $V < \SI{-2.5}{\volt}$ so that the quantum well is fully depleted underneath the gate, even underneath the aluminum strip (strong depletion).
The teal gates are normally set to moderately negative voltages (between $\SI{-2}{\volt}$ and $\SI{-1.5}{\volt}$) so that the quantum well is fully depleted underneath the gate, except underneath the aluminum strip (light blue), where the lowest sub-band is partially occupied (depletion). 
The white gates are set to voltages to form tunnel junctions and modulate the associated dot-nanowire or dot-dot tunnel coupling.
The red gates are set to more positive voltages (accumulation). 
\figpanel{b}~The corresponding gates can be seen in the SEM image, where we have also marked the locations of the quantum dots, the MZMs, the tunneling paths between them, and the interference loop that links them all.  
Note that in the measurements presented in the main text, helper gates HG1 and HG2 are set to depletion to reduce the required voltages on SC1 and SC2 to isolate the loop, though this voltage setting is not required.  
Additionally, in the main text where the focus is on the left loop, the dot gates corresponding to the middle loop (QC3 and QD4) are fully depleted and the remaining gates on the right loop are grounded.
}
\label{fig:device_configuration}
\end{figure*}

The qubit device shown in \Cref{fig:device_configuration} is a practical realization of the linear tetron of Refs.~\onlinecite{Fidkowski11b, Karzig17}.
The complete device has 7 quantum dots, three long ones and four small ones.
For concreteness, we call the long dots dot~2, dot~4, and dot~6; the small dots are dot~1, dot~3, dot~5, and dot~7. 
Dots 2 and 6 run parallel to the two topological sections of the nanowire, and dot~4 runs parallel to the middle trivial section of the nanowire.
Each quantum dot is covered by a plunger gate in the second gate layer, whose purpose is to set the electrical potential on the underlying dot.  We refer to the dot plunger gates using the convention ``QD\textit{i}''.
There are also ``cutter'' gates in the second gate layer. 
They cover the junctions: nanowire-dot~1, dot~1-dot~2, dot~2-dot~3, dot~3-nanowire, etc. as shown in \Cref{fig:device_configuration}a. 
Quantum dot cutter ``QC\textit{i}'' gates control inter-dot tunnel couplings; tunnel gates ``TG\textit{i}'' control coupling between the small dots and the nanowire; and source cutter ``SC\textit{i}'' gates control coupling between the small dots and the two-dimensional electron gas (2DEG) reservoirs.  
The 2DEG density in these reservoirs is set by the voltage on the helper gates ``HG\textit{i}'' which run from the dot region all the way to metallic Ohmic source contacts (denoted by purple boxes labeled ``S\textit{i}'' in \Cref{fig:device_configuration}a).
The gates QD1, QD3, QD5, and QD7 are used to de-tune dot states from the Fermi energy, thereby setting the effective tunneling amplitudes of the MZMs from the wire to dot~2, dot~4, and dot~6 via tunneling through the small dots. 
The lateral confinement of the dots is provided by the wire plunger ``WP\textit{i}'' gates and depletion gates ``DG\textit{i}''.

The two topological wire segments (segments under WP1 and WP3) are $\SI{2.96}{\micro\meter}$ long, and the trivial segment (segment under WP2) between them is $\SI{2.96}{\micro\meter}$ long.
Simulations of this device indicate that, in the topological phase, the coherence length has minimum value $\xi = \SI{1}{\micro\meter}$~\cite{Aghaee23}, so $e^{-L/\xi}\approx 0.05$.
The outer trivial segments (regions under side plungers SP1 and SP2) are $\SI{5}{\micro\meter}$ long; hence, the topological segments are far from the normal electrons at the Ohmic contacts at the ends of the Al strip (drain contacts D1 and D2).
The Al strip and, therefore, the nanowire beneath it is $\SI{60}{\nano\meter}$ wide, enabling full depletion of the InAs quantum well beneath the Al while providing sufficient screening of charged impurities in the dielectric.

The small dots (dots 1, 3, 5, 7) have a T-junction shape with a width of $\SI{200}{\nano\meter}$ where they meet the cutter gates and an effective length between the nanowire and dot~2 of $\SIrange{400}{500}{\nano\meter}$.
The long dots (dots 2, 4, 6) are $\SI{2.4}{\micro\meter}$ long.
The cutter gates are $\SI{100}{\nano\meter}$ wide. These lengths were chosen so that dot~2 could couple via dot~1 and dot~3 to the left topological segment's MZMs, which are $\approx \SI{3}{\micro\meter}$ apart; dots 1, 2, and 3 bridge this length.
By breaking the length up in this way, we ensure that all dots have level spacings $\gtrsim \SI{10}{\ueV}$ and dots 2, 4, and 6 have charging energy $\gtrsim \SI{45}{\ueV}$.
Meanwhile, dots 1, 3, 5, and 7 have charging energy $\gtrsim \SI{100}{\ueV}$.
Since the long dot level spacing is greater than the temperature, thermal smearing of the interference signal is minimized.

\subsection{Device fabrication}
\label{sec:fabrication}

The device presented here follows the same fabrication as dual-layer gate devices presented in the preceding work~\cite{Aghaee23}.
The \SI{6.5}{\nano \meter} thick Al strip on top of the semiconductor substrate features larger Al pads at each end of length \SI{1.5}{\micro\meter} and width \SI{3.5}{\micro\meter}, see ~\Cref{fig:device_configuration}a. The Al pads are Argon milled to ensure low-resistance Ohmic contact to electron beam lithography defined structures of Ti(\SI{10}{\nano \meter})/Au(\SI{140}{\nano \meter})/Ti(\SI{10}{\nano \meter}) by which the Al strip is grounded. Both types of contacts are normal in the typical operating regime.

The Ti(\SI{2}{\nano \meter})/Au(\SI{15}{\nano \meter})/Ti(\SI{2}{\nano \meter})/Al(\SI{10}{\nano \meter}) Gate layer 1 is separated from the semiconductor by Dielectric 1 which consists of $\approx \SI{2}{\nano \meter}$ AlO$_x$ and $\SI{10}{\nano \meter}$ HfO$_x$.
The electron beam lithography-defined gates in the first gate layer define the nanowire by depleting the 2DEG around the Al strip and also the quantum dots. 
The Ti(10nm)/Au(60nm) Gate layer 2 is separated from Gate 1 by Dielectric 2 which is $\SI{15}{\nano \meter}$ AlO$_x$. 
Layer 2 gates, also patterned by electron beam lithography, enable independent control of the coupling between the nanowire and quantum dots as well as the chemical potential of the quantum dots.

\subsection{Superconductor-semiconductor hybrid heterostructure design}
\label{sec:material-stack}

The heterostructure presented here is grown on commercially available semi-insulating InP wafers. 
A graded buffer is used to translate the lattice constant from that of InP, $\SI{0.587}{\nano \meter}$, to one near InAs, $\SI{0.605}{\nano \meter}$. 
The active region is constructed on the graded buffer and consists of a $\SI{25}{\nano \meter}$ In$_{0.845}$Al$_{0.155}$As lower barrier, a $\SI{9.1}{\nano \meter}$ thick InAs quantum well and a $\SI{6}{\nano \meter}$ thick In$_{0.88}$Al$_{0.12}$As top barrier, as well as the Al superconductor. 
The top barrier layer plays a critical role in fine-tuning the coupling between the superconductor and the 2DEG residing in the quantum well. 
For an Al parent gap which is $\Delta_0\sim\SI{290}{\micro\eV}$ in our device, we measure an induced gap $\DeltaInd\sim\SI{110}{\micro\eV}$. 
Another function of the top barrier layer is to separate the quantum well states from disorder on the dielectric-covered surface of the stack. 
From a Hall bar device simultaneously fabricated on the same chip as the nanowire device, we extract a peak mobility $\mu\sim \SI{75000}{\cm^2/\volt\second}$ at a density $n_\mathrm{max}\sim\SI{1.2e12}{\cm^{-2}}$.

Spin-orbit coupling strength in a hybrid superconductor-semiconductor structure is difficult to measure directly.
Using weak anti-localization measurements in shallow InAs 2DEGs, see, for example, Ref.~\onlinecite{Shabani16}, and typical values of the electric field (obtained from simulations assuming the band offset parameter measured in Ref.~\onlinecite{Schuwalow21}), we estimate that the Rashba spin-orbit coupling is in the range of $\SIrange{5}{10}{\milli\electronvolt\nm}$.

\subsection{Readout system}
\label{sec:readout_system}

\begin{figure}
\includegraphics[width=8cm]{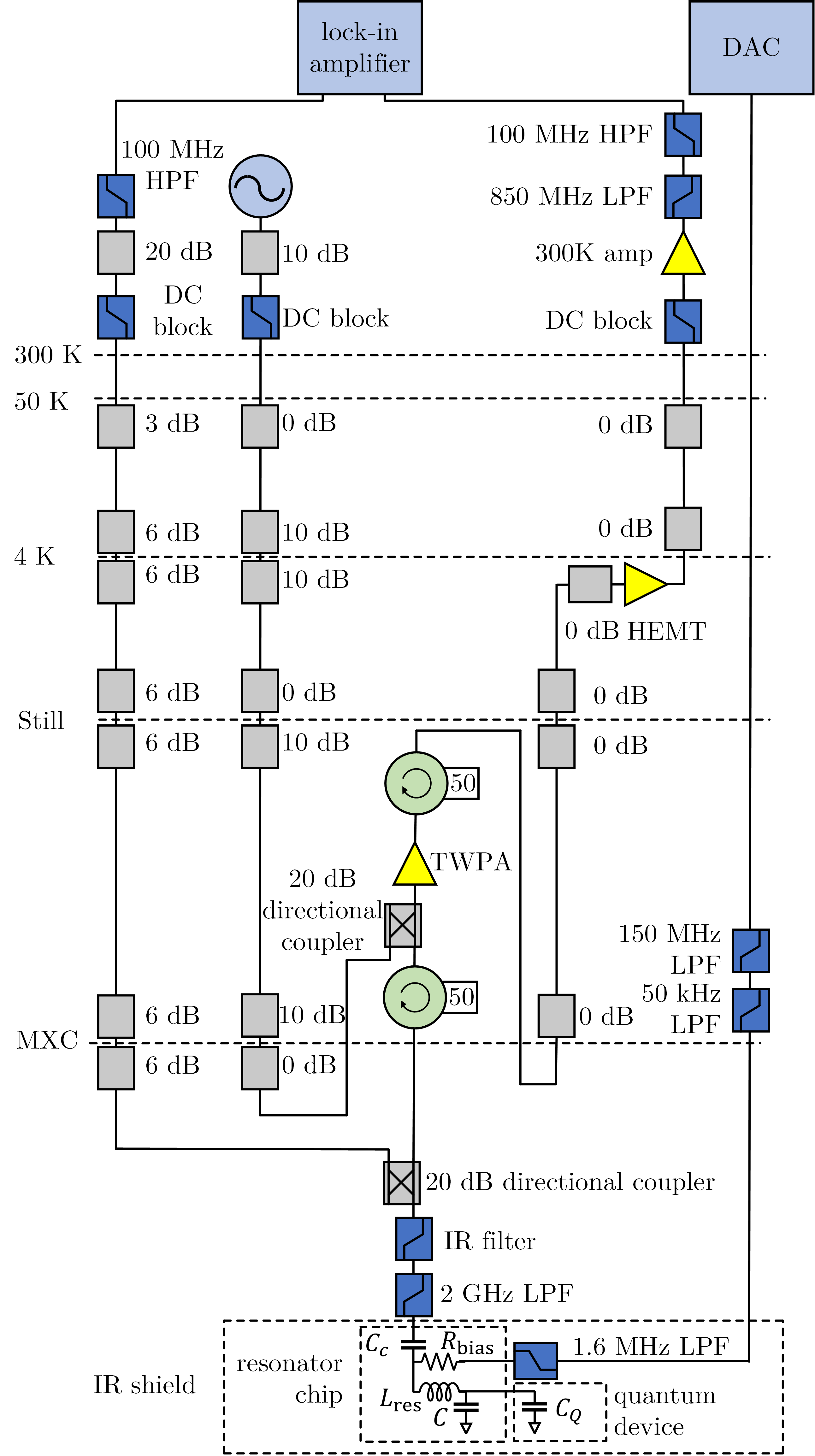}
\caption{
Schematic diagram of the reflectometry setup used for the measurement of fermion parity by dispersive gate sensing, and shielding applied to sample, and filtering of rf and dc lines. 
The resonator chip includes a bias tee for setting $\VQD{2}$, through $R_\mathrm{bias} = 16 \SI{}{\kilo\Omega}$, and rf signal is routed to the capacitively coupled port, through $C_\mathrm{c} \sim \SI{8}{\pico\farad}$, to the rest of the interferometry setup. 
Here, we have used the standard abbreviations for a digital-to-analog converter (DAC), low-pass filter (LPF), and high-pass filter (HPF).
}
\label{fig:readout_system}
\end{figure}

The microwave readout chain is shown in \Cref{fig:readout_system}. 
To measure fermion parity, the readout chain is designed to minimize the system noise temperature while maximizing the quasiparticle poisoning time $\tau_\mathrm{qpp}$.  
To this end, we use a near-quantum limited amplifier, the JTWPA~\cite{Macklin2015}, as the first amplifier in the reflectometry setup.  
At the same time, low-pass and IR filters and several layers of shielding are applied to suppress stray radiation impinging on the sample. 
A table of key parameters characterizing the readout system is given in \Cref{tab:simulation_parameters}. 

\begin{table}
\begin{center}
\begin{tabularx}{\columnwidth}{|c|C|C|W|C|W|C|C|c|}
\cline{1-8}
Run &
\shortstack[l]{$V_\mathrm{res}$ \\ $[\SI{}{\micro\volt}]$ } &
\shortstack[c]{$Q$ \\ \,} &
\shortstack[c]{$\kappa_\mathrm{ext}/2\pi$ \\
$[\SI{}{\mega\hertz}]$} &
\shortstack[c]{$\omega/2\pi$ \\
$[\SI{}{\mega\hertz}]$} &
\shortstack[c]{$C$ \\ 
$[\SI{}{\femto\farad}]$} &
\shortstack[c]{$T_\mathrm{N}$ \\
$[\SI{}{\kelvin}]$} &
\shortstack[c]{$\Delta \Cq$ \\
$[\SI{}{\atto\farad}]$}
\\ 
\cline{1-8}
A1 &
\qty{10\pm2}{} &
\qty{35\pm4}{} &
\qty{16\pm2}{} &
\qty{617}{} &
\qty{580\pm25}{} &
\qty{0.6\pm0.1}{} &
\qty{962}{} \\
B1 &
\qty{20\pm4}{} &
\qty{29\pm3}{} &
\qty{18\pm2}{} &
\qty{651}{} &
\qty{520\pm20}{} &
\qty{0.6\pm0.1}{} &
\qty{250}{} \\
\cline{1-8}
\end{tabularx}
\end{center}
\vskip -3mm
\caption{
Measured parameters for devices A and B (run 1), which are used to estimate the expected SNR.  
Both devices use an inductor with $L_\mathrm{res}$ = \SI{115}{\nano H}.
}
\label{tab:simulation_parameters}
\end{table}

\subsection{Software}

All measurements have been implemented using QCoDeS open-source python library for instrument control and data acquisition \cite{Nielsen24}.

\section{Theoretical model}
\label{sec:design-details}

\subsection{Basic features of the device design}
\label{sec:basic_concept}

In the topological superconducting phase of a nanowire, the low-energy physics is
characterized by two real fermionic
operators $\gamma_1$ and $\gamma_2$. These operators are localized at the left and right ends of the topological section, respectively, and satisfy the relationship $\{{\gamma_i}, {\gamma_j}\} = 2\delta_{ij}$.
The wire has two nearly-degenerate ground states with opposite fermion parities
$i\gamma_1 \gamma_2 = \pm 1$. The  splitting energy between these states, $\EM$, is approximately 
$\EM \sim \DeltaT \exp(-L/\xi)$ in the limit where $L\gg \xi$~\cite{Kitaev01}. Here, $\xi$ represents the coherence length, 
and $\DeltaT$ is the topological gap. This behavior holds also in the presence of disorder~\cite{Motrunich01}, although the coherence length $\xi$ is renormalized~\cite{Brouwer11b}. 

To measure the parity of Majorana zero modes (MZMs), we designed a device with three quantum dots (QDs) and a readout resonator. The middle QD (dot~2) is coupled to the resonator, while the side QDs (dot~1 and dot~3) act as tunable barriers. We show that the tunneling rate between dot~2 and the nanowire depends on the detuning energies in dot~1 and dot~3. We then derive a low-energy model for this system and estimate the quantum capacitance signal.

\subsection{Effective Hamiltonian for MZM-dot~1-dot~2 system}
\label{sec:triple-dot-effective-H}

We now develop an effective low-energy model for the MZM-dot~1-dot~2 system. This model facilitates the determination of the effective parameters in \Cref{eq:cq}, and it elucidates the distinction between balanced and unbalanced interferometers. The Hamiltonian for the MZM-dot~1-dot~2 system is given by
\begin{align}
H &\!=\!2\EM \left(c^\dag c\!-\!\frac{1}{2}\right)\!+\!\Delta_1\left(f^\dag f\!-\!\frac{1}{2}\right)\!+\!\Delta_2\left(d^\dag d\!-\!\frac{1}{2}\right)+H_t, \\
H_t &\!=\!\tm{1}f^\dag (c+c^\dag)+\tm{1}^*(c+c^\dag)f+ t_{12} f^\dag d+t_{12}^*d^\dag f. \nonumber
\end{align}
Here, we introduce non-local fermionic operators in the wire, denoted as $c=(\gamma_1+i \gamma_2)/2$ and $c^\dag = (\gamma_1-i \gamma_2) /2$. 
These operators encode the fermion occupation of a pair of Majorana zero modes, with Majorana splitting energy denoted by $\EM$. The operators $f^\dag$, $f$ and $d^\dag$, $d$ are fermion creation and annihilation operators in the small (e.g., dot~1)  and long (dot~2)  quantum dots, respectively. 
The detuning of quantum dots $i$ from the degeneracy point is given by $\Delta_{i}(\Ng{i})=\ECi{i} (1-2 \Ng{i})+\delta_i$. 
The parameters $\ECi{i}$, $\Ng{i}$, and $\delta_{i}$ denote the charging energy, dimensionless offset charge, and level spacing of the $i$th dot, respectively. 
The dimensionless offset charge is controlled by the gate voltage, expressed as $\Ng{i}=\alpha_i e \VQD{i}/2\ECi{i}$, where the lever arm $\alpha_i = 2 C_{\mathrm{g}i} \ECi{i}/e^2$. 
We assume that the level spacing in the quantum dots is large $\delta_i \gg \kB T$, $|t_{12}|$, $|\tm{1}|$ ($T$ being temperature), thereby enabling an effective description via a single-level approximation. 
Lastly, the matrix elements $\tm{1}$ and $t_{12}$ describe the tunneling between dot~1-MZM and dot~2-dot~1, respectively.   

\begin{widetext}
Let us first focus on the even parity sector of the combined MZM-dot~1-dot~2 system and derive the effective Hamiltonian when dot~1 is detuned from resonance. We use the basis convention $\ket{a,b,c,d} = \ket{0,1,0,1}_\mathrm{MZM}\otimes \ket{0,0,1,1}_\QD{1} \otimes \ket{0,1,1,0}_\QD{2}$. The corresponding Hamiltonian is given by
\begin{align}
H_\mathrm{even} =
\left(\begin{array}{cccc}
    - \EM - \Delta_+ & 0 & 0 & -t_m^* \\
    0 & \EM - \Delta_- & \tm{1}^*  & t_{12}^* \\
    0 & \tm{1} & - \EM + \Delta_+ & 0 \\
    - \tm{1} & t_{12} & 0 & \EM + \Delta_-
\end{array}\right)\!,
\end{align}
where $\Delta_\pm = (\Delta_1 \pm \Delta_2)/2$.
The above Hamiltonian can be simplified in the limit of large dot~1 detuning. 
Assuming that $|\Delta_1| \gg |\Delta_2|$, $|\tm{1}|$, $|t_{12}|$, $\EM$ we can perform a Schrieffer-Wolff transformation and ``integrate out'' dot~1 to arrive at
\begin{align}
\!\! {\tilde H}_\mathrm{even} =
\left(\begin{array}{cccc}
    - \EM - \Delta_+ -\cfrac{|\tm{1}|^2}{\Delta_1} & \cfrac{t_{12} \tm{1}^*}{\Delta_1} & 0 & 0 \\
    \cfrac{t^*_1 \tm{1}}{\Delta_1} & \EM - \Delta_- -\cfrac{|\tm{1}|^2 + |t_{12}|^2}{\Delta_1} & 0  & 0 \\
    0 & 0  &  - \EM + \Delta_+ + \cfrac{|\tm{1}|^2}{\Delta_1} & \cfrac{t_{12}^* \tm{1}}{\Delta_1} \\
    0 & 0  & \cfrac{t_{12} \tm{1}^*}{\Delta_1}  &  \EM + \Delta_- + \cfrac{|\tm{1}|^2+|t_{12}|^2}{\Delta_1}
\end{array}\right)\!.
\end{align}
In this derivation, we have disregarded small terms of $O(1/|\Delta_1|^2)$. 
The upper left block is representative of an empty dot~1 (i.e., when $\Delta_1>0$), while the lower right block is for an occupied dot~1 (i.e., when $\Delta_1<0$). 
The off-diagonal terms within each sub-block correspond to co-tunneling through dot~1. 
Consequently, the side dots essentially facilitate a tunable MZM-dot~2 coupling. Furthermore, tunneling between dot~2-dot~1 and dot~1-MZM leads to a correction of detuning energies. 
It's important to note that these corrections, induced by quantum fluctuations, differ for empty and occupied dot~2 states. 
Consequently, as shown below, the peak of the quantum capacitance experiences a shift with respect to the $\Delta_2 = 0$ line even when $\EM = 0$.
\end{widetext}

An analogous analysis can be performed for the odd parity sector of the combined MZM-dot~1-dot~2 subsystem. 
In the appropriate basis, the effective Hamiltonian ${\tilde H}_\mathrm{odd}$ can be derived from ${\tilde H}_\mathrm{even}$ by substituting $\EM$ with $-\EM$. 
The quantum capacitance for a state $n$ with energy $\varepsilon_n$ can be expressed as $\Cq^{(n)} = -e^2 \alpha^2 \,{\partial^2 {\varepsilon_n}}/{\partial\ED^2}$. 
Assuming thermal occupation, the quantum capacitance for even and odd parity sectors reads:
\begin{multline}
\Cq(Z) =
\frac{e^2 \alpha^2 2|t_\mathrm{eff}|^2 }
{\left[(\ED + Z 2\EM)^{2} +
4|t_\mathrm{eff}|^2\right]^{3/2}}\times \\
\tanh \left(\frac{\sqrt{(\ED + Z2 \EM)^{2} + 4|t_\mathrm{eff}|^2}}{2 \kB T}\right),
\label{eqn:C_Q-one-dot}
\end{multline}
where $\ED = \Delta_2 - |t_{12}|^2 / |\Delta_1|$, $t_\mathrm{eff} = t_{12}\tm{1}^* / \Delta_1$ and $Z=\pm 1$ for even/odd parity states of the combined MZM-dot~1-dot~2 subsystem.
The temperature dependence is due to thermal occupation of the excited state within each parity sector, which suppresses $\Cq$.
In the interferometer setup, when dot~2 is coupled to both MZMs, one obtains \Cref{eq:cq}, where the effective parameters are given by the expressions above.

\subsection{Dynamical response of the TQDI within linear response theory}
\label{sec:linear-response}

In this section, we calculate the dynamical response of the TQDI coupled to a readout resonator. By extending the effective model derived earlier to the MZM-dot~1-dot~2-dot~3-MZM setup, we arrive at the following effective Hamiltonian:
\begin{multline}
\HF = 
\ED d^\dag d 
+ 2\EM \left(c^\dag c - \frac{1}{2}\right) \\
+ \tR d^\dag (c + c^\dag)+\tR^*(c + c^\dag)d \\
+i\tL d^\dag (c^\dag-c)
+ i\tL^*(c^\dag-c)d.
\label{eq:HF}
\end{multline}
Here $\ED = \Delta_2 - |t_{12}|^2 / |\Delta_1| - |t_{23}|^2 / |\Delta_3|$,  $\tL = t_{12}\tm{1}^* / \Delta_1$, and $\tR = t_{23}\tm{2}^* / \Delta_3$. 
The quantities $\Delta_{i}(\Ng{i}) = E_{Ci} (1-2 \Ng{i})+\delta_i$ are the detunings of the quantum dots from their degeneracy points, as defined above. 
The couplings $\tm{1}$ and $\tm{2}$ denote QD-MZM couplings for dot~1 and dot~3, respectively, while $t_{12}$, $t_{23}$ describe tunneling between dot~1-dot~2 and dot~2-dot~3, respectively.

To perform a readout of the fermion parity split between $\gamma_1$ and $\gamma_2$, we couple dot~2 to a readout resonator with an angular frequency $\omega_\mathrm{r}$. 
In the rotating frame, the full Hamiltonian, including the microwave photons, is given by
\begin{equation}
     H = \HF + (\omega_\mathrm{r}-\omega) \hat{a}^\dagger \hat{a} + g d^\dag d (\hat{a}e^{-i\omega t}+\hat{a}^\dagger e^{i\omega t}).
     \label{eq:H_with_photons}
\end{equation}
Here  $a^\dag$ and $a$ denote the photon creation and annihilation operators in the resonator, and the coupling to the resonator is defined as $g=\alpha_2 e \sqrt{\hbar \omega_\mathrm{r}/2C}$ where $C$ is the total effective capacitance of the readout circuit, which includes the capacitance of the resonator itself.   

To understand the impact of a quantum device on the resonator, it is necessary to calculate both the frequency shift and the system-induced loss on the photon degrees of freedom. 
This can be achieved by integrating out the fermionic degrees of freedom. 
Upon performing this procedure and retaining only the lowest-order corrections in $g$~--- an approach equivalent to linear response theory~--- one finds that the resonator’s response is encapsulated in the following correlation function:
\begin{multline}
K^{R}(\omega) 
= -i\frac{g^2}{2} \int\frac{d\omega_1}{2\pi} \,
\mathrm{Tr} \Bigl[\hat{V}G^K(\omega_1)\hat{V}G^R(\omega_1 + \omega)
\\
+ \hat{V}G^A(\omega_1)\hat{V}G^K(\omega_1 + \omega)\Bigr].
\label{eq:KO}
\end{multline}
Here $G^R(\omega)$, $G^A(\omega)$ and $G^K(\omega)$ denote the retarded, advanced and Keldysh Green's functions, respectively, for the Hamiltonian $\HF$. The vertex operators $\hat{V}$ are defined as $\hat{V} =\partial [G^R(\omega)]^{-1} / \partial\ED$. The real and imaginary components of $K^{R}(\omega)$ represent the shift in the resonator frequency and photon loss, respectively. 
The response of the system can be interpreted as a dynamical quantum capacitive shift $\Cq(\omega)$ relative to the capacitance of the readout circuit $C$ which we model by an effective lumped element circuit (see \Cref{fig:minimal_model}c and \Cref{fig:readout_system}). Expanding for $\Cq(\omega)\ll C$ yields the relation 
\begin{align}
\frac{\Cq(\omega)}{2C }=-\frac{K^{R}(\omega)}{\omega_\mathrm{r}}.
\end{align}
As $\omega$ and $T$ approach zero, the expression for $\Cq(\omega)$ recovers the quantum capacitance of a ground state of the system. Note that in the low-frequency limit there is in general an additional contribution to the capacitance from incoherent tunneling processes, see, e.g. Ref.~\onlinecite{Esterli19}. Throughout this paper we focus on the quantum capacitance which dominates the response for $\tC \gtrsim \kB T$ and for probe frequencies fast compared to thermalization time scales.  

Finally, by projecting the Hamiltonian \Cref{eq:HF} onto both the even and odd parity sectors of the QD-wire system and utilizing \Cref{eq:KO}, we can derive an expression for the dynamical quantum capacitance of the TQDI within each parity sector: 
\begin{multline}
\Cq(Z,\phi,\omega) 
= \frac{2 e^2 \alpha^2 |\tC(Z,\phi)|^2}
{\sqrt{(\ED + Z 2\EM)^{2} + 4|\tC(Z,\phi)|^2}}
\\
\times \frac{\tanh \left(\sqrt{(\ED + Z 2\EM)^2 + 4|\tC(Z,\phi)|^2} / 2\kB T\right)}
{(\ED + Z 2\EM)^2 + 4|\tC(Z,\phi)|^2 - (\omega +i \eta)^2},
\label{eq:CqZ}
\end{multline}
where $\eta$ is the broadening of the resonance due to the coupling to the environment (see also the discussion below). The term $\tC(Z,\phi)$ in \Cref{eq:CqZ} is given by
\begin{align}
\left|\tC(Z,\phi)\right|^2 = |\tL|^2 + |\tR|^2 +
2 Z |\tL| |\tR|  \sin\phi. \label{eq:tCphi}
\end{align}
Here, $\phi$ is the phase difference between the left and right QD-wire junctions, which is controlled by the external magnetic flux, given by $\phi=2\pi \Phi/\Phi_0+\phi_0$ with $\Phi_0 = h/e$. $\phi_0$ is a flux-independent offset. $\Cq$ for each parity sector exhibits periodicity in flux with period $h/e$. Both the real and imaginary parts of $\Cq(Z,\phi,\omega)$ contain significant information regarding microscopic parameters,
which can be used in comparing
simulation to experiment.

The difference in the dynamical quantum capacitance between the even and odd parity sectors is given by
\begin{align}
\Delta \Cq(\phi, \omega)= \left|\Cq(1,\phi,\omega)-\Cq(-1,\phi,\omega)\right|.
\end{align}
In the static limit $\omega \rightarrow 0$, one recovers \Cref{eq:delta_cq} of the main text.
Using \Cref{eq:CqZ}, we estimate that the maximum value of $|\Delta \Cq|$ is approximately $\SI{1}{\femto\farad}$
using the parameters $\EM = \SI{1}{\micro\eV}$,
$\tR = \SI{3}{\micro\eV}$,
$\tR = \SI{3}{\micro\eV}$,
$\eta = \SI{1}{\micro\eV}$,
$\kB T= \SI{4}{\micro\eV}$,
$\omega = \SI{2}{\micro\eV}$,
$E_{C2} = \SI{45}{\micro\eV}$,
$\alpha = 0.5$. As discussed in the main text (see \Cref{sec:parity_measurement}), the peaks in $\Delta \Cq(\phi, \omega)$ for even and odd parities occur at $\ED = -2\EM$ and $\ED = 2\EM$, respectively. As $\EM$ increases, at $\ED = -2\EM$, $\Cq$
is suppressed in the odd parity sector
and enhanced in the even parity sector,
and vice versa at $\ED = 2\EM$.


\subsection{Dynamical $\Cq$ calculation using open system dynamics} 
\label{sec:dynamical-cq}

The linear response theory outlined above provides an intuition for how the dynamical $\Cq$ response arises in our setup. For quantitative comparison to the experimental data, it is important to capture larger drive amplitudes beyond linear response. We therefore use a different approach that is based on simulating the dynamics of the system in an open-system framework. 
We first note that in typical dispersive gate sensing setups, the resonator photons are well-described by the classical limit $\hat{a}\to a$ of a large number of photons. 
This can be used to reduce \Cref{eq:H_with_photons}, or its generalization including all 3 QDs explicitly, to a time-dependent problem in the fermionic Hilbert space. 
Specifically, the coupling to the (classical) photon fields can be captured via
\begin{equation}
\tHF = 
\HF + 2g|a| d^\dagger d \cos(\omega t).
\label{eq:HF_with_drive}
\end{equation}
The equation of motion of the classical field $a$ takes the form
\begin{equation}
\dot{a} = 
i[\omega-\omega_\mathrm{r}-K(\omega)]a -\frac{\kappa}{2}a-\sqrt{\kappa_e}b_\mathrm{in},
\label{eq:aEOM}
\end{equation}
where $K(\omega) = g \langle d^\dagger d \rangle (t) e^{i\omega t}/|a|$ and we included photon decay $\kappa=\kappa_\mathrm{ext}+\kappa_\mathrm{int}$ due to internal loss $\kappa_\mathrm{int}$ and coupling to the readout line $\kappa_\mathrm{ext}$ and the external drive $b_\mathrm{in}$ via standard input-output theory, see e.g. Ref.~\onlinecite{Clerk10}.

We then solve the equations \Cref{eq:HF_with_drive,eq:aEOM} including the effects of a noisy environment. We introduce the coupled system-bath Hamiltonian
\begin{equation}
H = 
\tHF + H_\mathrm{sb},
\end{equation}
where $H_\mathrm{sb} = \sum_i X_i \Phi_i + H_\mathrm{env}$ contains both the Hamiltonian for the environment and the coupling between system and environment. 
Here, $X_i$ are a set of system operators and $\Phi_i$ are bath operators. 
In general, further approximations must be made to solve such complex Hamiltonians. Here, we work within the Born-Markov approximation using the Universal Lindblad master equation (ULE) approach described in Ref.~\onlinecite{Nathan20}. 
Within this framework, the environment is described entirely via the spectral functions of the bath operators,
\begin{equation}
S_i(\omega) = 
\int dt\, e^{i \omega t} \langle \Phi_i(0) \Phi_i(t) \rangle.
\label{eqn:bsf}
\end{equation}
We adopt a notation where the system operators $X_i(t)$ are dimensionless, and all dimensionful prefactors are absorbed into the definition of $\Phi_i$.
By appropriately choosing $X_i(t)$ and $S_i(\omega)$, we can capture charge noise, phonon noise and other noise sources. For the purpose of simulating a system that is driven by the resonator, it is essential to capture the coupling to a low-temperature environment which leads to a steady state that is close to thermal equilibrium. 
The condition that the environment is in thermal equilibrium can be stated in terms of the spectral functions as $S_i(\omega)/S_i(-\omega)=\exp(\hbar \omega/ \kB T)$.

To connect to \Cref{eq:aEOM}, we set $\langle d^\dagger d\rangle(t)=\tr\{\rho(t) d^\dagger d\}$ with the density matrix $\rho(t)$ obtained from numerically solving the ULE. This approach allows us to numerically simulate the dynamical $\Cq$ response of the MZM-dot~1-dot~2-dot~3-MZM system including the backaction from the finite drive amplitude. The dissipative terms in the ULE formalism naturally lead to  the imaginary part of the response captured by the phenomenological parameter $\eta$ in the linear response theory \Cref{eq:CqZ}. 
For a more detailed description of this approach to determining the rf response of the system as well as a detailed comparison to the linear response regime, see Ref.~\onlinecite{rf_response}.

An important contributor to the detuning noise of QDs is $1/f$ charge noise which is particularly strong at low frequencies. Due to the long-time correlations of this noise it cannot easily be described via the spectral functions of the ULE framework. Instead, we capture this noise by making use of the ability to handle arbitrary time dependence of the Hamiltonian $\tHF$ and treat the low-frequency charge noise explicitly as classical noise on the corresponding terms in the Hamiltonian. Considering the example of noise on the dimensionless gate charge $\Ng{i}$, we consider an ensemble of noise trajectories that satisfies the autocorrelation function
\begin{equation}
\langle \Ng{i}(0) \Ng{i}(t) \rangle 
= \int \frac{d\omega}{2\pi}\, e^{-i \omega t} S^c_{i}(\omega),
\label{eqn:autocorr}
\end{equation}
where we have used the superscript $c$ to indicate the difference to the bath correlation function in \Cref{eqn:bsf}. A similar approach has been used previously, e.g. in Ref.~\onlinecite{Mishmash20}, and we follow the approach discussed there to numerically draw noise realizations from this ensemble. In general, it may be necessary to introduce low- and high-frequency cutoffs, which we will discuss below in \Cref{sec:toy-model-sims}.

\subsection{Numerical details}
\label{sec:toy-model-sims}

\begin{figure}
\includegraphics[width=\columnwidth]{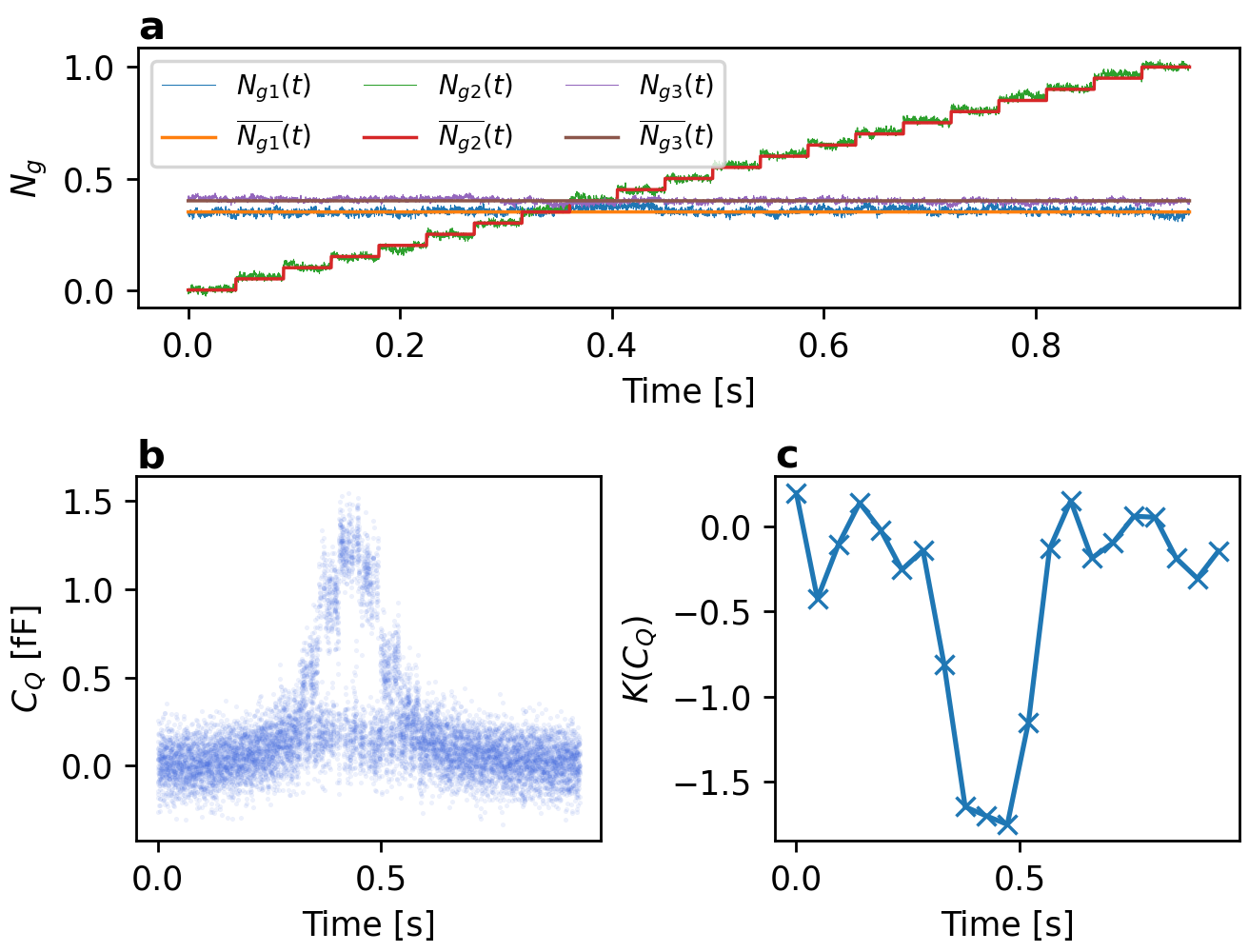}
\vspace{-6.5mm}
\caption{
Illustration of simulations of parity readout maps. The example shown here corresponds to a cut at fixed flux $\Phi = 0.5 h/2e$.
\figpanel{a}~Evolution of the gate voltages $\Ng{i}(t)$, where $\barNg{i}(t)$ indicates the intended evolution for each gate, including the stepping of the voltage on QD2 by 0.01 every $\SI{45}{\milli\second}$ and keeping dot~1 fixed at $\Ng{1} = 0.35$ and dot~3 at $\Ng{3} = 0.4$. 
$\Ng{i}(t)$ indicates the noisy trajectory obtained after including the effects of low-frequency charge noise.
\figpanel{b}~$\Cq(t)$ obtained at each point in the time trajectory.
\figpanel{c}~Kurtosis of the timetraces at each $\Ng{2}$.
}
\label{fig:mpr-example}
\end{figure}

We now provide additional details on how our simulations are performed and, in particular, how classical noise is treated to closely match the experimental measurements.


The starting point is the model of \Cref{sec:basic_concept}, including the two Majorana zero modes $\gamma_{1,2}$ with splitting energy $\EM$ and three quantum dots (dots 1 through 3), tuned via gate charges $\Ng{1}$ through $\Ng{3}$. The charging energies in the isolated QD regime are given in \Cref{fig:dot_tuneup}. When the interferometer loop is formed, the charging energies are renormalized down due to quantum charge fluctuations involving virtual transitions to states of higher energy [outside the low-energy subspace defined in \Cref{eq:HF}]. 
To estimate this effect, one can consider a model including an additional level in dot~2 (quantum dot with the smallest level spacing) and obtain a low-energy model using a Schrieffer-Wolff transformation, similarly to the derivation of \Cref{sec:triple-dot-effective-H}. By expanding the derived effective model near the charge degeneracy, one can estimate the renormalization of model parameters due to virtual transitions to excited states. Using $|t_{12}|=|t_{23}|=\SI{12}{\ueV}$ and extracted bare charging energies, one finds the renormalized charging energies to be approximately given $\tilde{E}_{C1} \approx \SI{140}{\ueV}$, $\tilde{E}_{C2}\approx\SI{45}{\ueV}$, and $\tilde{E}_{C3}\approx\SI{100}{\ueV}$, respectively. The simulations presented in the main body of the paper employ these renormalized charging energies.

We treat the evolution of this system in the ULE formalism of \Cref{sec:dynamical-cq}, where the coupling to the environment is included through charge noise on each of the three quantum dots. The corresponding Lindblad operators are obtained within the ULE framework for system operators $X_i = n_\QD{i}$, $i=1,2,3$, where $n_\QD{i}$ is the number operator for each of the three dots, and the spectral function is given by
\begin{equation} \label{eq:S_g}
S_\mathrm{g}(\omega) 
= \frac{\hbar^2 \gamma}{1 + \exp(-\hbar\omega / \kB T)}
\end{equation}
with $\gamma = \SI{1}{\giga\hertz}$ and $T = \SI{50}{mK}$. This choice of $\gamma$ is motivated by extrapolating the strength of $1/\omega$ charge noise measured in \Cref{sec:charge_noise} to $\hbar\omega \approx \SI{5}{\ueV}$, which is the relevant energy scale for these simulations. It is worth noting that the precise value chosen here has only very small effect on the real part of $\Cq$, as it mainly controls the rate of thermalization.
We ignore noise on the couplings, which is expected to be much weaker than charge noise on the quantum dot detunings \cite{Khindanov21a}. We introduce a drive on $\Ng{2}$, i.e. we replace $\Ng{2} \rightarrow \Ng{2} + (A_\mathrm{rf}/2\ECi{2}) \cos(\omega t)$, where we choose $A_\mathrm{rf} = \SI{5}{\micro\electronvolt}$ (note that $A_\mathrm{rf} = \alpha e V_\mathrm{res}$, where  $\alpha \approx 0.5$ is the lever arm) and $\ECi{2}$ corresponds to the charging energy of dot~2. This corresponds to \Cref{eq:HF_with_drive} with
constant $|a|$. This thus defines a simulated response $\Cq(\Ng{1}, \Ng{2}, \Ng{3}, \Phi, Z)$, where $\Phi$ is the flux enclosed in the loop and $Z$ is the overall parity of the system.

The final step is to include low-frequency charge noise, readout noise, parity flips, and the way the experimental data is acquired. As an example, we can consider the parity readout maps as shown in, e.g., \Cref{fig:deviceA1_parity_measurements}a. These consist of timetraces of about $\SI{50}{\milli\second}$ duration at each point in the $(\Ng{2}, B_\perp)$ plane, where $\Ng{2}$ constitutes the inner loop and $B_\perp$ the outer loop. Measurements are taken at approximately 100 points in $\Ng{2}$ (typically spanning 5 charge transitions) and 70 points in $B_\perp$; the total time for a single measurement panel is thus approximately $\SI{320}{\second}$. To generate this behavior, we generate time traces of all parameters of the system. For each gate voltage, we generate a trajectory $\Ng{i}(t) = \barNg{i}(t) + \delta\Ng{i}(t)$, where $\barNg{i}(t)$ is a function describing the ideal parameter evolution and $\delta \Ng{i}(t)$ is a noise trajectory whose power spectral density (cf \Cref{eqn:autocorr}) is given by $S^c(\omega) = \alpha_C^2/\omega$ (see \Cref{fig:mpr-example}a). Here, we choose $\alpha_C = 0.00675$, which corresponds to $\sqrt{S_0} = \SI{1.35}{\micro\electronvolt}$ in units of detuning on QD2, consistent with the measured charge noise on our devices (see \Cref{sec:charge_noise}). Furthermore, we take the inverse of the integration time as high-frequency cutoff and the total duration of the experiment as natural low-frequency cutoff. The tuning parameter evolution $\barNg{i}(t)$ is a constant in the case of the side dots, whereas for dot~2 it is a function that increases from 0 to 1 in 20 steps over $20 \times \SI{50}{\milli\second}$. For the parity $Z(t)$, we generate a telegraph noise trajectory with a symmetric transition rate of $\Gamma_\mathrm{qpp} = (\SI{1}{\milli\second})^{-1}$. Finally, the flux $\Phi(t)$ is stepped from 0 to 4 $h/2e$ in 70 increments. The readout noise is added as Gaussian noise with standard deviation $\sigma(\Cq)$, where we choose $\sigma(\Cq) = \SI{0.105}{\femto\farad}$ in line with the experimental observation in \Cref{fig:deviceA1_parity_measurements}. Taken together, this allows us to define $\Cq(t) = \Cq(\Ng{1}(t), \Ng{2}(t), \Ng{3}(t), \Phi(t), Z(t))$ in a data format that exactly reproduces that of the experimental data, see \Cref{fig:mpr-example}b,c.

\subsection{Expected signal-to-noise ratio}
\label{sec:expected-SNR}

To estimate the expected Signal-to-Noise ratio (SNR) of the measurement of $\Delta \Cq$ we consider the steady state solution of \Cref{eq:aEOM}
\begin{equation}
\label{eqn:input-output}
a = \frac{\sqrt{\kappa_\mathrm{ext}}b_\mathrm{in}}{i(\omega-\omega_Z) - i\kappa/2}
\end{equation}
where $\omega_Z=\omega_\mathrm{r}+K(\omega,Z)$. Within standard input-output theory, the relation between the incoming and outgoing photon fields is given by $b_\mathrm{out}=b_\mathrm{in}+\sqrt{\kappa_\mathrm{ext}}a$~\cite{Clerk10}. The relevant signal $S=|b_\mathrm{out,+1}-b_\mathrm{out,-1}|/2$ is then given by the shift of the output fields relative to each other. The typical frequency shift $|\omega_{+1}-\omega_{-1}|$ in our devices is small compared to $\kappa$ which allows for an expansion of the signal to lowest order in the capacitive shift, 
\begin{equation}
    S =  \frac{\sqrt{\kappa_\mathrm{ext}}|a| \omega_\mathrm{r}}{2\kappa C} \Delta \Cq\,
\end{equation}
where we used the regime of optimal drive frequency $\omega = (\omega_{+1}+\omega_{-1})/2$. Recasting this in terms of the voltage amplitude $V_\mathrm{res}=\sqrt{2\hbar \omega_\mathrm{r}/C}|a|$ in the resonator and adding noise from an amplification chain of effective noise temperature $T_\mathrm{N}$, the signal to noise ratio is given by
\begin{equation}
    \mathrm{SNR} = V_\mathrm{res}\frac{Q}{2}\sqrt{\frac{\kappa_\mathrm{ext} \tau_\mathrm{m}}{C k_\mathrm{B} T_\mathrm{N}}} \Delta \Cq\,
    \label{eq:SNR}
\end{equation}
where $Q=\omega_\mathrm{r}/\kappa$ and $\tau_\mathrm{m}$ is the integration time of the measurement.

Using \Cref{eq:SNR}, we can estimate the typical SNR in our setup using 
the parameters in \Cref{tab:simulation_parameters}.  These parameters are obtained with measurements on the same device shown in \Cref{fig:deviceA1_parity_measurements}. We find $\text{SNR}=0.8(2)$ in $\tau_\mathrm{m}=\SI{1}{\micro\second}$.
\Cref{eq:SNR} appears to yield an SNR that is monotonically increasing with $V_\mathrm{res}$. However, for some sufficiently large $V_\mathrm{res}$, the SNR becomes limited by backaction. As $V_\mathrm{res}$ increases, power broadening will decrease the dynamical $\Delta \Cq$ which yields an optimum in the drive amplitude which is typically of the order $e\alpha V_\mathrm{res}\sim(2\tC-\hbar \omega)$. Note that the dynamical $\Cq$ formalism outlined above can capture this backaction via  \Cref{eq:HF_with_drive} and \Cref{eq:aEOM}.

\section{Analysis of the quasi-MZM scenario}
\label{sec:quasi-MZM}

\begin{figure}
\centering
{\hspace{-1.3cm}\includegraphics[width=0.6\columnwidth]{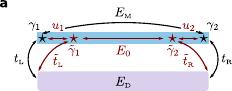}} \\
\includegraphics[width=0.9\columnwidth]{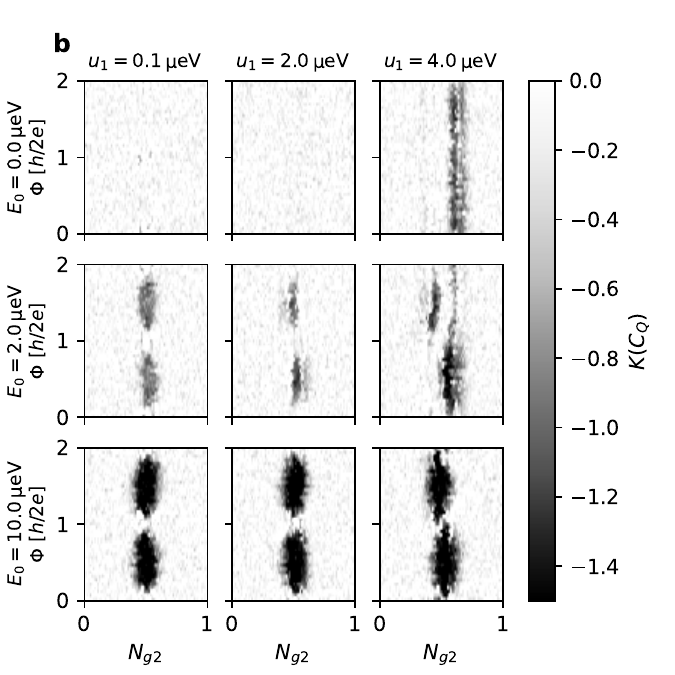}
\vspace{-3mm}
\caption{
\figpanel{a}~Schematic representation of the various terms in the Hamiltonian of \Cref{eq:HF_qMZM}. 
The terms in addition to the 2MZM model are highlighted in red.
\figpanel{b}~Simulated response in the case of a wire with 4 Majorana modes and 3QDs. The corresponding mapping to the single QD scenario is described in \Cref{sec:triple-dot-effective-H}. 
From top to bottom rows $E_0$ is increased. 
From left to right columns $u_1$ is increased. 
Other parameters: $\tm{1}=\tm{2}=\SI{5}{\micro\electronvolt}$, $\tilde{t}_\mathrm{m1}=\tilde{t}_\mathrm{m2}=0$, $t_{12}=t_{23}=\SI{10}{\micro\electronvolt}$, $\Ng{1}=1-\Ng{2}=0.35$, $\EM=0$, $u_2=0.7u_1$, $T=\SI{50}{\milli\kelvin}$, drive amplitude $A_\mathrm{rf}=\SI{4.5}{\micro\electronvolt}$.
}
\label{fig:qmzm}
\end{figure}

In this section, we consider the situation
in which there are two additional hidden Majorana modes, in addition to the end MZMs that couple to QD1 and QD3.
We begin with the effective Hamiltonian
\begin{align}
\label{eq:HF_qMZM}
& H_\mathrm{qMZM} = \HF+\HF^{(1)}, \\
& \HF^{(1)} = 
iu_1 \gamma_1 {\tilde \gamma}_1
+ iu_2 \gamma_2 {\tilde \gamma}_2
+ i{E_0} {\tilde \gamma}_1{\tilde \gamma}_2\nonumber \\
& \qquad\quad + ({\tilde t}^{}_\mathrm{\scriptscriptstyle L} d^\dagger - {\tilde t}^*_\mathrm{\scriptscriptstyle L} d) {\tilde \gamma}_1
+ ({\tilde t}^{}_\mathrm{\scriptscriptstyle R} d^\dagger - {\tilde t}^*_\mathrm{\scriptscriptstyle R} d) {\tilde \gamma}_2,
\end{align}
where $\HF$ was defined in \Cref{eq:HF}. 
Here, ${\tilde \gamma}_{1,2}$ are the hidden Majorana modes which are somewhere in the middle of the wire, with ${\tilde \gamma}_{i}$ closer to $\gamma_{i}$ to which it tunnels with amplitude $u_i$. 
Meanwhile the tunneling amplitude between ${\tilde \gamma}_{1}$ and ${\tilde \gamma}_{2}$ is $E_0$. 
A schematic representation of this model is depicted in \Cref{fig:qmzm}a.

In the quasi-MZM scenario, $\tilde{\gamma}_{1,2}$ are localized at their respective ends of the wire, leading the presence of two trivial low-energy states with small but finite splitting $u_1,u_2$ while the across-wire coupling of these local states is small, $E_0\approx 0$. To build intuition we thus consider the model in \Cref{eq:HF_qMZM} where  $|\tL|$, $|\tR| \gg |u_1|$, $|u_2|$. We take $|\tilde{t}_\mathrm{\scriptscriptstyle L}|$, $|\tilde{t}_\mathrm{\scriptscriptstyle R}|$, $E_0 \to 0$ for simplicity, and find an analytical solution for this limit. We then give numerical results for the more general case. We first project the Hamiltonian in \Cref{eq:HF_qMZM} to the fixed parity subspaces and perform a Schrieffer-Wolff transformation to find  
\begin{widetext}
\begin{align}\label{eq:HqMZMP}
H_\mathrm{qMZM}^{(e/o)}\approx\left(
  \begin{array}{cccc}
    -E_{\mathrm{\scriptscriptstyle M1}} P -\delta \ED & t_\mathrm{\scriptscriptstyle L1}+t_\mathrm{\scriptscriptstyle R1} P & 0 & 0 \\
t^*_\mathrm{\scriptscriptstyle L1}+t^*_\mathrm{\scriptscriptstyle R1} P & \ED+E_\mathrm{\scriptscriptstyle M1} P - \delta \ED & 0 & 0 \\
    0 & 0 & E_\mathrm{\scriptscriptstyle M1} P + \delta \ED  & t_\mathrm{\scriptscriptstyle L1}-t_\mathrm{\scriptscriptstyle R1} P \\
    0 & 0 & t^*_\mathrm{\scriptscriptstyle L1}-t^*_\mathrm{\scriptscriptstyle R1} P & \ED - E_\mathrm{\scriptscriptstyle M1} P + \delta \ED \\
  \end{array}
\right).
\end{align}
\end{widetext}
Using the notation $|a,b,c\rangle\equiv |a\rangle_\QD{} \otimes |b\rangle_\mathrm{MZM} \otimes |c\rangle_\mathrm{\tilde MZM}$, the basis used above is defined as $\{|000\rangle$, $|110\rangle$, $|011\rangle$, $|101\rangle\}$ in the even sector ($P=1$) and $\{|010\rangle$, $|100\rangle$, $|001\rangle$, $|111\rangle\}$ in the odd sector ($P = -1$).

The effective parameters in \Cref{eq:HqMZMP} are defined as $\delta \ED = \ED u_1 u_2 / D(\phi)$, $E_\mathrm{\scriptscriptstyle M1}=\EM+\ED(u_1^2+u_2^2)/2D(\phi)$, $t_\mathrm{\scriptscriptstyle R1} = -i \tR+e^{i\phi}u_2^2 \tL/D(\phi)$, $t_\mathrm{\scriptscriptstyle L1}=e^{i\phi}\tL-i u_1^2 \tR/D(\phi)$, $D(\phi)=\EM \ED-2 \tL \tR \sin \phi$. Perturbation theory, which yields the Hamiltonian \Cref{eq:HqMZMP}, is valid as long as $|u_1|$, $|u_2|\ll \sqrt{|D(\phi)|}$. The higher-order terms omitted in \Cref{eq:HqMZMP} will generally lead to transitions within each parity sector that occur faster than both our measurement time and the parity switching time due to non-equilibrium quasiparticles. Thus, we assume that our measurements sample the equilibrium density matrix within each parity sector.          

The (static) quantum capacitance for each parity sector can now be calculated using the Hamiltonian in \Cref{eq:HqMZMP}, as was previously discussed in \Cref{eqn:C_Q-one-dot}. Examining \Cref{eq:HqMZMP}, we notice that each block $i\tilde{\gamma}_1\tilde{\gamma}_2=\pm 1$ shows a similar capacitive response as is obtained in the 2-MZM scenario. Crucially though, now both $Z=\pm 1$ sectors contribute at fixed overall parity $P$, where $P$ is the total parity of all four MZMs and the dot while $Z$ is the parity of $\gamma_{1,2}$ and the dot. This leads to a cancellation of $\Delta \Cq^\mathrm{qMZM}=\Cq^\mathrm{qMZM}(P=+1)-\Cq^\mathrm{qMZM}(P=-1)$ up to small corrections when $\delta \ED$ is finite. 
Indeed, one can show that
\begin{multline}
\!\! \Delta \Cq^\mathrm{qMZM} = 
-\left[ \Cq(Z=+1)-\Cq(Z=-1) \right] \\
\times \delta \ED \, \frac{1}{\kB T} \frac{
    4 \cosh \cfrac{E^{(+)}}{2\kB T} \cosh \cfrac{E^{(-)}}{2\kB T}
}{\left(
    \cosh \cfrac{E^{(+)}}{2\kB T} + \cosh \cfrac{E^{(-)}}{2\kB T}
\right)^2}, \!\!
\end{multline}
where $E^{(Z)}=\sqrt{(\ED\pm 2E_\mathrm{\scriptscriptstyle M1})^2+4 |\tC(Z)|^2}$ and $\Cq$ is the corresponding expression for the 2MZM scenario in \Cref{eq:cq} with the model parameters renormalized via $\tL\to t_\mathrm{\scriptscriptstyle L1}$, $\tR \to t_\mathrm{\scriptscriptstyle R1}$ and $\EM\to E_\mathrm{\scriptscriptstyle M1}$.

The second line is a product of three factors; the first vanishes for $\delta\ED = 0$ and the product of the second
and third vanishes for $T=0$. Hence, $\Delta \Cq^\mathrm{qMZM}$ vanishes if either of these conditions holds. More generally,
$\Delta \Cq^\mathrm{qMZM}$ is suppressed by at least a factor of $\delta\ED/\kB T$ relative to the topological 2-MZM scenario. Note that for $\ED,\tL,\tR\sim t$, $u_1,u_2\sim u$, and at optimal flux, $\delta \ED/\kB T\sim u^2/2t\kB T$, which  would lead to a suppression of $\Delta \Cq$ by at least a factor of two for $u<\SI{3}{\micro \electronvolt}$ for the parameters extracted in \Cref{fig:qd_mzm}. For low temperatures relative to $E^{(+)}-E^{(-)}$ there is an additional suppression $\propto \exp(-|E^{(+)}-E^{(-)}|/2\kB T)$. Given this suppression, we generally expect $\Delta \Cq^\mathrm{qMZM}$ to be significantly smaller in the quasi-MZM scenario than in the topological regime, which we verify below via detailed
simulation.


We now study the quasi-MZM model \Cref{eq:HF_qMZM} numerically, explicitly including all 3 QDs. Using the same methods as described in \Cref{sec:dynamical-cq} and \Cref{sec:toy-model-sims} we show the dependence of the capacitive response as a function of $E_0$ and the local splitting $u_1,u_2$ in \Cref{fig:qmzm}.  Here we set $\EM=0$ in order to attempt to reproduce the $h/2e$ periodic signal in the 4-Majorana-mode scenario. Moreover, we set the couplings of the Majorana modes $\tilde{\gamma}_1$ and $\tilde{\gamma}_2$ to their respective QDs to zero for simplicity, which doesn't affect the observed qualitative behavior.

When $E_0$ is large, ${\tilde \gamma}_{1,2}$ are effectively frozen out and can be ignored. This could occur, for instance, if there were a small region of trivial phase in the middle of the wire, with Majorana modes ${\tilde \gamma}_{1,2}$ at the ends of this region, far from the
ends of the wire. Hence, this case is actually the case of \Cref{eq:HF}, rather than quasi-MZMs. In the simulations
presented in \Cref{fig:qmzm}, we indeed observe that for sizable $E_0$ we recover the expected $h/2e$ periodic signal of the topological regime. For small $E_0$, we see only a weak signal for small $u_1$ in line with the analytical model discussed above. For larger $u_1$ there is a sizable $\Delta \Cq$ signal but without a clear flux dependence. This scenario is similar to scenario (c) in \Cref{sec:interferometry_regimes} and describes the bimodal response of the QD coupled to a local state and the absence of interference \cite{Hell18}. Intermediate values of $E_0$ interpolate between these scenarios but clear $h/2e$ flux periodicity is only observed for $E_0 \gtrsim u_1,u_2$. The $\Delta \Cq$ response saturates to the topological scenario in this regime when $E_0$ exceeds $\kB T$.

We thus conclude that in order to reproduce the observed $h/2e$ periodic response with large values of $\Delta \Cq$ in a 4-Majorana model in general requires $E_0 \gg u_1,u_2$ and $E_0>\kB T$, which is better understood in terms of the model of \Cref{eq:HF} with an additional subgap state at energy $2E_0$ rather than a non-topological quasi-MZM scenario.




\section{Data Analysis}
\label{sec:data-analysis}

\subsection{$\Cq$ conversion}
\label{sec:cq_conversion}

The microwave signal collected in dispersive gate sensing measurement encodes the microwave susceptibility of the quantum device under study~\cite{Park20, Metzger21, Kurilovich21, Fatemi22}.
To make connection with theory, we explicitly convert the measured microwave signal into a quantum capacitance $\Cq$.  
Previously, $\Cq$ conversion has been performed with methods based on resonator fitting~\cite{Malinowski22a, Malinowski22b}.  
These can yield both the real and the imaginary parts of the quantum capacitance (also called the quantum conductance)~\cite{Malinowski22b}.  
Importantly, with the aid of a reference measurement composed of a frequency scan of the readout resonator, they allow conversion of a single IQ pair into a complex quantum capacitance.  

Here we use a similar technique, which leverages symmetries and a small parameter expansion to obtain an analytical approximation for the mapping between the microwave signal and quantum capacitance.  
As in Refs.~\onlinecite{Malinowski22a, Malinowski22b}, the first step in the conversion process is to determine the change in the resonance frequency $\delta \omega_\mathrm{r}$ and the loss rate $\delta \kappa_\mathrm{int}$ relative to the reference trace.

\begin{figure}
\includegraphics[width=5.5cm]{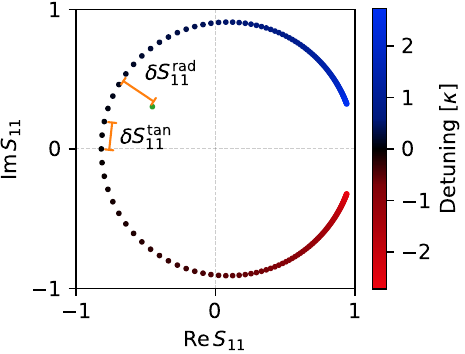}
\vskip -2mm
\caption{
Geometrical depiction of the $\Cq$ conversion process for an IQ pair with $\delta \omega_\mathrm{r} = \kappa/8$, $\delta \kappa_\mathrm{int} = \kappa/5$. 
The circle shows the reflection of an overcoupled resonator in IQ space. 
The green disc represents a sample IQ pair to be converted into a complex $\Cq$. 
The real (imaginary) component of $\Cq$ encodes $\delta \omega_\mathrm{r}$ ($\delta \kappa_\mathrm{int}$) and causes translations along the arc (radius) of the circle.
}
\label{fig:cq_conversion}
\end{figure}

We first describe how the change in the resonance frequency $\delta \omega_\mathrm{r}$ is detected.  
After correction of electrical delay, the microwave reflection $S_{11}$ of a readout resonator with resonant frequency $\omega_\mathrm{r}$, internal loss rate $\kappa_\mathrm{int}$, external loss rate $\kappa_\mathrm{ext}$, and total linewidth $\kappa = \kappa_\mathrm{int} + \kappa_\mathrm{ext}$ is
\begin{equation}
S_{11} = 1 - \frac{2 \kappa_\mathrm{ext}/\kappa} {1+ 2i\left(\omega-\omega_\mathrm{r}\right)/\kappa},
\label{eq:S11}
\end{equation}
when it is probed at a frequency $\omega$.
Crucially, \Cref{eq:S11} depends on $\omega$ and $\omega_\mathrm{r}$ only through their difference $\omega-\omega_\mathrm{r}$.  
This implies that the change in $S_{11}$ from a small shift in $\omega_\mathrm{r}$ is identical to the change resulting from an equal and opposite detuning of the probe frequency $\omega$.  
(Formally the shift in $\omega_\mathrm{r}$ needs not be small, but in practice non-idealities like the background ripple in the reflectometry setup or uncorrected electrical delay break the symmetry on which this argument rests.)  
The reference trace of $S_{11}$ as function of the probe frequency $\omega$ provides a look-up table of the reflection coefficients measured at different detunings.  
To determine the shift in the resonance frequency, we identify the point on the reference trace nearest to the IQ pair to be converted (which is probed at frequency $\omega_\mathrm{r}$). 
The detuning of this point from the resonance frequency in the reference trace is equal and opposite to the desired frequency shift $\delta \omega_\mathrm{r}$. 
This process is illustrated in \Cref{fig:cq_conversion}. 
The green disc indicates the IQ pair to be converted. 
An orange line of length $\delta S_{11}^\mathrm{rad}$ connects it to the nearest point on the circle, which has a known detuning that can be read from the colorbar.

To determine the change in the loss rate $\kappa_\mathrm{int}$ we leverage the fact that both $\delta\omega_\mathrm{r}/\kappa$, $\delta \kappa_\mathrm{int}/\kappa \ll 1$. 
By expanding \Cref{eq:S11} to first order in these small parameters, we determine their effect on the reflection coefficient:
\begin{equation}
\delta S_{11} \approx 2\frac{\kappa_\mathrm{ext}}{\kappa} \left(\frac{\delta \kappa_\mathrm{int}}{\kappa} - 2i\frac{\delta \omega_\mathrm{r}}{\kappa} \right).
\label{eq:S11_exp}
\end{equation}
In \Cref{eq:S11_exp}, $\delta \kappa_\mathrm{int}$ changes the real part of $S_{11}$. 
Geometrically this corresponds to a radial translation toward or away from the circle that the reference trace forms in the IQ plane. 
For this reason we denote the distance between the circle of the reference trace and the IQ pair to be converted as $\delta S_{11}^\mathrm{rad}$ (see \Cref{fig:cq_conversion}).
As with the determination of $\delta \omega_\mathrm{r}$, here again the reference trace can be used as a look-up table to read off the translation in IQ space $\delta S_{11}^\mathrm{tan}$ from a small (relative to $\kappa$) detuning $\delta \omega$.  
(We denote the translation with the superscript ``tan'' because it is tangential to the circle of the reference trace.) 
The change in the loss rate is then
\begin{equation}
\delta \kappa_\mathrm{int} \approx 2 \frac{\delta \omega}{\delta S_{11}^\mathrm{tan}} \delta S_{11}^\mathrm{rad}.
\end{equation}
In practice, selection of $\delta \omega$ involves a tradeoff~--- it should be chosen to be as large as possible (to reduce inaccuracies from readout noise) while also being much less than the linewidth $\kappa$ (to reduce inaccuracies from the series expansion).  
We choose $\delta \omega \approx \kappa/20$. 

Finally, to convert the computed complex frequency shifts into a complex quantum capacitance $\Cq$ we use the fact that $\Cq$ is small relative to the total capacitance $C$.  
Expanding $\omega_r = 1/\sqrt{L_\mathrm{res}\left(C + \Cq\right)}$ we have
\begin{equation}
\label{eq:tilde-CQ-def}
\frac{K}{\omega_\mathrm{r}} = -\frac{1}{2} \frac{\Cq}{C} -
\frac{1}{2} \frac{\delta L(\Bperp)}{L_\mathrm{res}} +
\mathcal{O}\left( \frac{\Cq}{C}\right)^2.
\end{equation}
Here $K = \delta{\omega}_\mathrm{r} - i\delta\kappa_\mathrm{int}/2$ denotes the complex response of the system (see also \Cref{sec:design-details}), the real part of which encodes the shift in the resonance frequency. We have introduced the quantity
$\delta L(\Bperp)$ to denote the dependence of the resonator's inductance $L_\mathrm{res}$ on $\Bperp$ 
which can shift
the resonator response over and above any $\Cq$-dependent
shift that occurs.
We define ${\tilde C}_\mathrm{\scriptscriptstyle Q} \equiv \Cq
+ C \,\delta L(\Bperp)/L$ which includes this effect.
We have chosen to use the same reference trace for all values of the out-of-plane field $B_\perp$, which may lead to a slow, systematic change of the extracted ${\tilde C}_\mathrm{\scriptscriptstyle Q}$ over the field range. 
This can be seen, e.g., in \Cref{fig:deviceA1_parity_measurements}b in the small overall curvature of the lower $\tCq$ branch. Indeed, ${\tilde C}_\mathrm{\scriptscriptstyle Q}$
can be negative.
To determine $\Cq$ itself, we would need
to independently measure $\delta L(\Bperp)$
(e.g. in a Coulomb valley, where $\Cq=0$)
and subtract its contribution, but the
effect of $\delta L(\Bperp)$
cancels out of $\Delta\Cq$ and the kurtosis $K(\Cq)$. 

Following convention, the imaginary part encodes  $-\kappa_\mathrm{int}/2$, such that a positive Im $\Cq$ corresponds to an increase in $\kappa_\mathrm{int}$.  We therefore have
\begin{eqnarray}
\frac{\mathcal{R}\left[\Cq\right]}{C} &=& -2 \frac{\delta \omega_\mathrm{r}}{\omega_\mathrm{r}}, \nonumber \\
\frac{\mathcal{I}\left[\Cq\right]}{C} &=& \frac{\delta \kappa_\mathrm{int}}{\omega_\mathrm{r}}.
\end{eqnarray}
The capacitance $C$ is computed with knowledge of the inductance $L_\mathrm{res}$ on the resonator chip and resonance frequency, $C \approx \left( \omega_\mathrm{r}^2 L_\mathrm{res}\right)^{-1}$.

We conclude this section with a practical note: measuring the IQ pair to be converted with a probe frequency $\omega \neq \omega_\mathrm{r}$ leads to conversion inaccuracy from the microwave background non-idealities. To remedy that, the IQ data can be transformed before performing the $\Cq$ conversion procedure. This transformation starts by rotating around the center of the resonance circle in IQ space by $\arg S_{11}(\omega_\mathrm{r}) - \arg S_{11}(\omega)$. The center of the circle can be determined by fitting an arc near resonance using an algebraic fit, such as Pratt's method~\cite{Probst2015}. The data is then scaled by the ratio of point densities ${\delta\omega}/{\delta S_{11}^\mathrm{tan}}$ at $\omega_\mathrm{r}$ and $\omega$ to account for the frequency-dependent phasal density of IQ pairs.

\subsection{Kurtosis}
\label{sec:kurtosis}

The kurtosis is defined as
\begin{equation} \label{eqn:kurtosis}
K(\Cq) = \frac{\mu_4(\Cq)}{[\mu_2(\Cq)]^2} - 3.
\end{equation}
Here, $\mu_n(\Cq) = N^{-1} \sum_{i=1}^N [\Cq(t_i) - \bar{C}_\Q]^n$
is the $n$th central moment of the timetrace $\Cq(t_i)$ with $N$ points and $\bar{C}_\Q$ its mean. The kurtosis is a good dimensionless measure of the how broad a random distribution is compared to a Gaussian. Here, it is of particular interest because it distinguishes between a Gaussian distribution, for which it takes the value $K=0$, and a bimodal distribution of two well-separated Gaussians, for which it takes a value $K < 0$. Other standard tests for distinguishing unimodal from multimodal distributions
(such as Pearson's criterion and the bimodality
coefficient) rely on functions of the kurtosis and skewness.

\subsection{Estimated SNR}
\label{sec:estimated-snr}

Recall that, in \Cref{sec:expected-SNR}, we estimated an expected $\text{SNR}=0.8(2)$ in $\tau_\mathrm{m}=\SI{1}{\micro\second}$ for device A based on our theoretical models.
We now discuss the SNR that we achieved in the time trace depicted in \Cref{fig:deviceA1_parity_measurements}(f).
Panel \Cref{fig:deviceA1_parity_measurements}g shows the measured data in the complex $\Cq$ plane. 
Note that we applied an overall shift to center $\text{Im}\, \Cq$ of the data around zero.
Focusing on the distribution of $\text{Re}\, \Cq$ shown in panel h, the distributions are centered around $\bar{C}_{\Q,+} = \SI{321}{\atto\farad}$ and $\bar{C}_{\Q,-} = \SI{1283}{\atto\farad}$, with standard deviations $\sigma_1 = \SI{101}{\atto\farad}$ and $\sigma_2 = \SI{92}{\atto\farad}$.
The SNR is given by $|\bar{C}_{\Q,+} - \bar{C}_{\Q,-}| / (\sigma_1 + \sigma_2) = 5.0$ for an effective integration time of $\SI{91}{\micro\second}$. The slight difference in standard deviation as well as height of the Gaussians can be attributed to the finite sample size used for fitting.
We find $\text{SNR}=0.52$ in $\tau_\mathrm{m}=\SI{1}{\micro\second}$, which is comparable to the estimated SNR value and allows $\text{SNR}=1$ to be obtained in $\tau_\mathrm{m}=\SI{3.7}{\micro\second}$ of integration time. 
We conclude that the sizable capacitive shifts enable fast single-shot readout even with relatively low-$Q$ resonators.
Our result compares favorably with the single-shot readout of gate-based spin qubits with off-chip resonators demonstrated in Refs.~\onlinecite{Pakkiam18, West19, Urdampilleta18}.

In the context of measurement-based quantum computation, an important performance metric is the probability of assignment errors, which needs to take into account the probability of state flips during the measurement. To this end, we define this quantity as
\begin{equation}
p_\mathrm{err} = 
\frac{1}{2} \biggl[ 1 - 
\exp\Bigl(-\frac{\tau_\mathrm{m}}{\tau_\mathrm{qpp}}\Bigr) 
\erf\Bigl(\frac{\SNR(\tau_\mathrm{m})}{\sqrt{2}}\Bigr) 
\biggr].
\end{equation}
Assuming $\tau_\mathrm{qpp}=\SI{2}{\milli\second}$ and the SNR extracted above, we find an optimal measurement time of $\SI{32.5}{\micro\second}$ and a corresponding $p_\mathrm{err}\approx1\%$. We note that this is the optimal measurement time at optimal flux, whereas for other flux values a longer measurement time is better able to distinguish the parity sectors. For this reason, we have used a longer measurement time throughout this paper.

\section{Device tune-up}
\label{sec:device-tune-up}

This section describes the general procedure to tune up the left TQDI loop in order to perform the interferometry measurements described in the main text.  A similar procedure can be used to tune up the right TQDI loop.  

Device tune-up requires both dc transport and rf dispersive gate sensing measurements. We first perform coarse tuning of quantum dots separated from the wire to the appropriate configuration satisfying the requirements described in \Cref{sec:scales}. In this step, each individual dot is formed separately and the rf drive is calibrated. After tuning the dots individually, the triple quantum dot system is tuned via dispersive gate sensing. 
As a next step, we use the dc transport measurements that comprise the topological gap protocol (TGP)~\cite{Aghaee23} to tune the nanowire into the relevant parameter regime in terms of wire plunger and in-plane magnetic field where we expect the topological phase. Finally, we proceed to tune the interferometer to optimize the $\Delta \Cq$ signal. 

\begin{figure*}
\includegraphics[width=17.8cm]{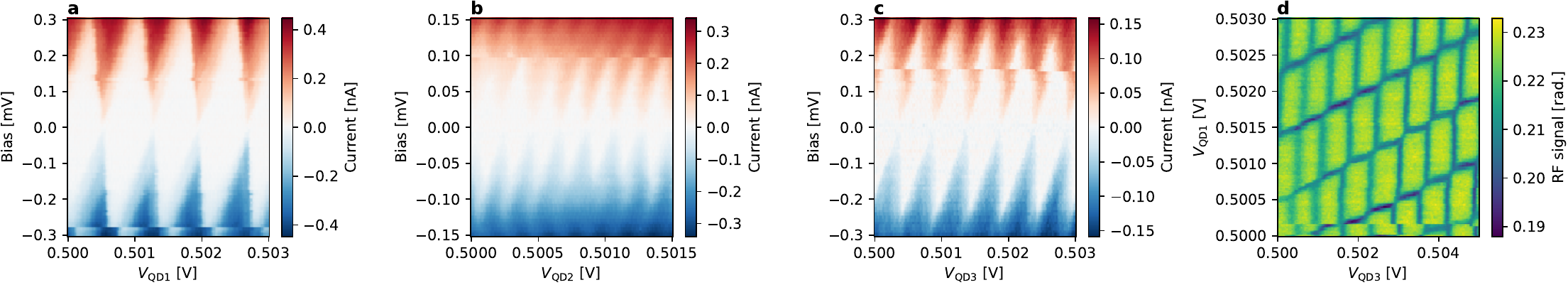}
\vskip -2mm
\caption{
Representative dot tune-up data. 
\figpanel{a-c}~Transport measurements of the respective single quantum dots when fully isolated from the nanowire (TG1 and TG2 set below their respective threshold voltages). 
\figpanel{d}~Dispersive gate sensing measurement of the charge stability diagram of the fully isolated triple-dot.  
TG1, TG2, SC1, and SC2 are all set below their respective threshold voltages. 
In this regime, the charging energies that have been extracted are as follows (averaged over the measured charge states) $E_{C1} = \SI{180}{\micro\electronvolt}$, $E_{C2} = \SI{60}{\micro\electronvolt}$, and $E_{C3} = \SI{130}{\micro\electronvolt}$. 
We note that our procedure for extracting the charging energies has $\approx 10\%$ uncertainty.
}
\label{fig:dot_tuneup}
\end{figure*}

\subsection{Pre-TGP wire transport}
First, the Ohmic contacts S1 and S2 are isolated from one another and the other Ohmic contacts by setting depletion gates DG1, DG2, and DG3 and cutters QC1 and QC2 below their respective threshold voltages.  In this configuration, the depletion voltage, induced gap, and parent gap of the topological wire segment are measured using local and non-local conductance spectroscopy as described in Ref.~\onlinecite{Aghaee23}.  

\subsection{Dot pre-tuning}
\label{sec:dot-tuning}

Next, rough tuneup of the quantum dots is done via dc transport. 
We first isolate the dots from the wire by setting TG1 and TG2 below threshold.  Transport is measured from Ohmic contact S1 to S2 with each dot formed by setting the adjacent cutter gates (i.e. SC or QC gate) to the tunneling regime and dot plungers and the remaining cutter gates in the current path to accumulation.  
A map of Coulomb diamonds as shown in \Cref{fig:dot_tuneup} is used to verify successful tune-up. These maps can also be used to extract the lever arms and bare charging energies of the quantum dots as well as the level spacing. We extract charging energies of $\SI{180}{\micro \electronvolt}$, $\SI{60}{\micro \electronvolt}$, and $\SI{130}{\micro \electronvolt}$ for dot~1, dot~2, and dot~3, respectively. From \Cref{fig:dot_tuneup}a we estimate the level spacing of the left quantum dot $\approx \SI{100}{\micro \electronvolt}$. Extrapolating to the $\approx 5$ times larger dot~2 (see  \Cref{fig:device_design}b)  yields a level spacing of $\approx \SI{20}{\micro \electronvolt}$ which indeed significantly exceeds the temperature as assumed in our theoretical modeling.

Finally, the conductance is measured while sweeping rf drive power and $\VQD{2}$ to observe broadening of the Coulomb blockade peaks and extract the lever arm used to convert rf drive power to voltage reaching the device.

We subsequently form a fully isolated triple quantum dot dot~1-dot~2-dot~3 by setting SC1 and SC2 below their threshold voltages while also leaving TG1 and TG2 below threshold. The dot plungers are left at the optimal value found in the single dot tune-up steps. First, a double quantum dot (DQD) is formed by setting QC2 below threshold.  The rf response of dot~2 is monitored while sweeping the voltages $V_\mathrm{QC1}$ and $V_\mathrm{QD1}$.  The inter-dot quantum capacitance will give rise to a measurable rf response in a narrow window of QC1 voltage.  Below this range the rf response is suppressed as the tunnel coupling drops below the temperature. Above this range, the magnitude of the quantum capacitance is suppressed by the large anti-crossing (and subsequently small curvature) of the DQD ground state.  Once this optimal range in QC1 is identified, QC1 is set below threshold and the same procedure is repeated for QC2.  With the optimal ranges of the two QC gates identified, a triple dot is formed in this isolated configuration by setting both QC gates to their optimal values.  A triple dot charge stability diagram like the one shown in \Cref{fig:dot_tuneup}d is used to verify successful tuneup of the triple dot.

\subsection{Topological gap protocol}

With the optimal voltages for the dots identified, we now proceed with running the TGP.  SC1 and SC2 are set back to accumulation while TG1 and TG2 are set to the tunneling regime to enable transport measurements on the wire again. The dot plungers are unchanged from the preceding step, and QC1 and QC2 are set below their threshold voltages to isolate S1 from S2.  With the auxiliary gates configured in this manner, stages 1 and 2 of the TGP are run as described in Ref.~\onlinecite{Aghaee23}.

\subsection{Tuning the TQDI loop}
\label{sec:MPR-tuneup}
After completion of a successful TGP, we select an in-plane magnetic field and WP1 voltage range with both sizeable transport gap and ZBP's at both ends of the wire.  With this field and WP1 voltage selected, TG1 and TG2 are varied to achieve a strong coupling to the ZBP's.  Once this coupling is established, SC1 and SC2 are set below threshold and all remaining measurements are done with dispersive gate sensing.
In order to establish a loop configuration, we set QC1 and QC2 back to their optimal values found in the earlier triple dot tuning step. QD-MZM coupling and interferometry measurements are then performed as described in the main text.

\subsection{QD-MZM tuning}

\begin{figure}
\includegraphics[width=\columnwidth]{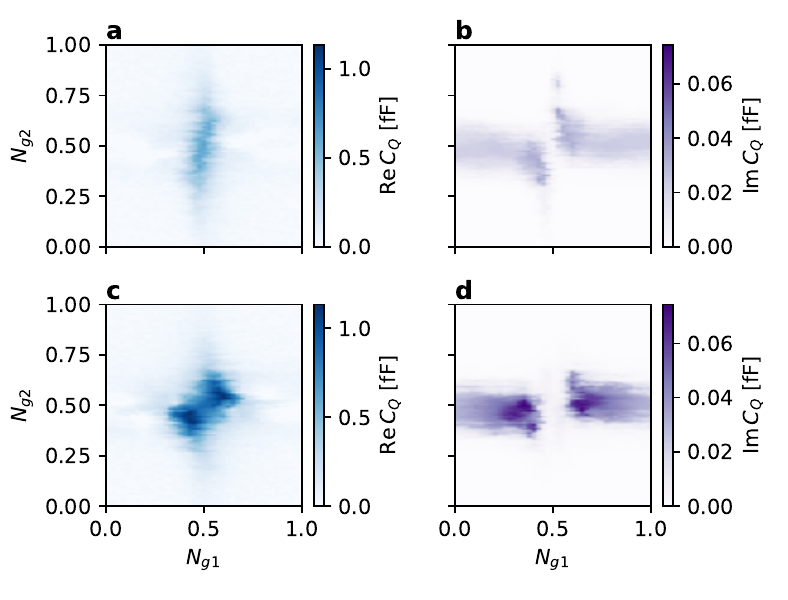}
\vspace{-8mm}
\caption{
Simulations of the configuration used for extracting the QD-MZM couplings, where one of the side dots (here dot~3) is maximally detuned while the gate voltages on the other two dots are being varied. 
The couplings used here are $t_{12} = t_{23} = \SI{12}{\micro\eV}$, and $\tm{2} = \SI{6}{\micro\eV}$ with $\tm{1} = \SI{2}{\micro\eV}$ in the top panels and $\tm{1} = \SI{6}{\micro\eV}$ in the bottom panels; left and right panels show the real and imaginary part of the $\Cq$ response, respectively. 
Unlike the parity measurement, the length of time traces is chosen to be $\SI{10}{\milli\second}$.}
\label{fig:qd_mzm_sim}
\end{figure}

The same approach discussed in \Cref{sec:toy-model-sims} can be used to simulate the response in the QD-MZM configuration, where one of the side quantum dots is completely detuned so that the coupling between the dot~2 and the topological wire is dominated by the other QD-MZM couplings. \Cref{fig:qd_mzm_sim} shows the $\Cq$ response of dot~2 simulated for this configuration in the limit of very small and very large coupling between $\gamma_1$ and dot~1, while dot~3 is fully detuned. One can see that both the shape and magnitude of the $\Cq$ response depend significantly on $\tm{1}$.
As discussed in the main text, by comparing the measured results quantitatively to features of these maps, we can identify parameters for the model introduced in \Cref{sec:basic_concept} that best describe the tuning configuration of the device.

\section{Measurement reproducibility and cross-checks}
\label{sec:reproducibility-and-checks}

In this section, we review an additional measurement performed on device A and a measurement of another device, which we call device B, to demonstrate the reproducibility of the measurements presented in the main text across measurement runs and devices. We also discuss additional cross-checks to further support our main conclusions.

\subsection{Second measurement of device A and measurement of device B}
\label{sec:measurement-A2-B}

\begin{figure*}
\includegraphics[width=17.9cm]{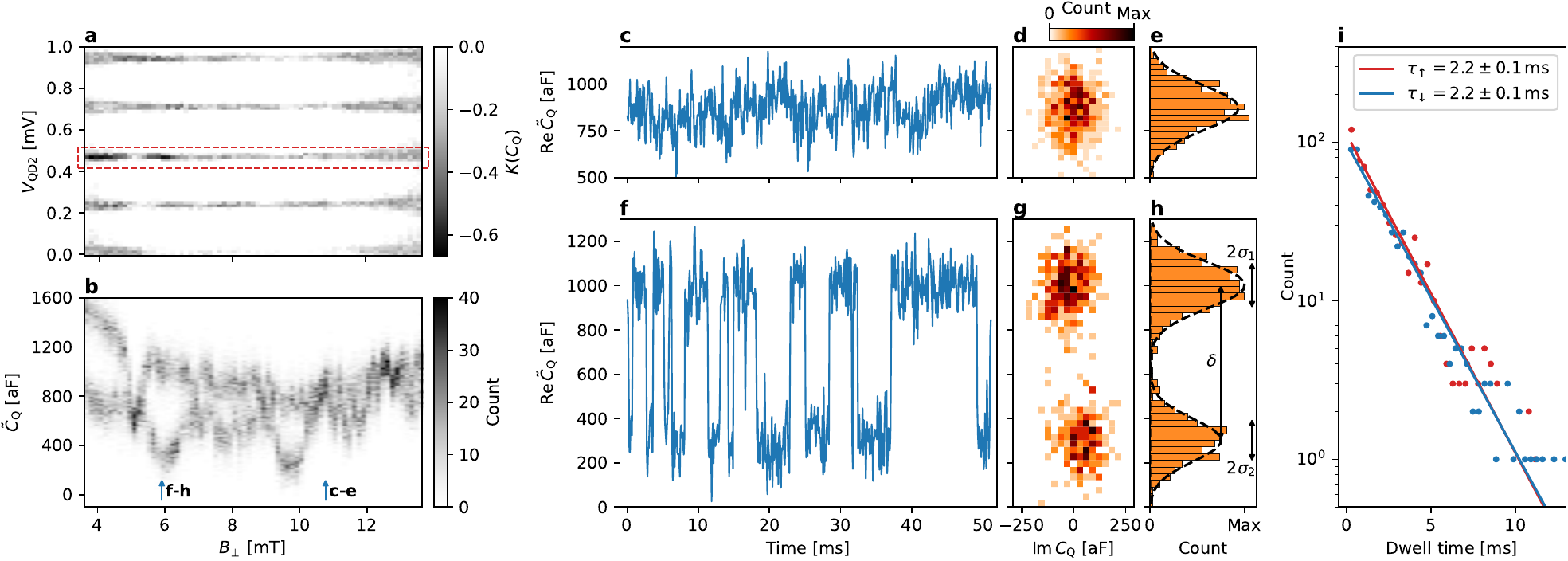}
\vspace{-5mm}
\caption{
Measurement A2: a second measurement of device A.
\figpanel{a}~Kurtosis in the measured quantum capacitance $K(\Cq)$ of dot~2 in device A as a function of $B_\perp$ and $\VQD{2}$ following the tune-up procedure of \Cref{sec:device-tune-up}. 
\figpanel{b}~A histogram of $\tCq$ values as a function of flux for the $\VQD{2}$ value in the middle of the dashed red rectangle in panel a, showing clear bimodality that is flux-dependent with period $h/2e$.  
\figpanel{c,f}~Time traces at the two flux values marked by the vertical arrows in panel b, corresponding to minimal (panel c) and maximal (panel f) $\Delta \Cq$.
\figpanel{d,g}~The raw rf signal converted to complex $\tCq$ by the method described in \Cref{sec:cq_conversion} for the time trace shown in panels c and f. 
\figpanel{e,h}~Histograms of $\mathrm{Re}\,\tCq$ with Gaussian fits for an extraction of the $\mathrm{SNR} = 3.50$.
\figpanel{i}~A histogram of dwell times aggregated over all values of $B_\perp$ where the signal shows bimodality. Fitting to an exponential shows that the up and down dwell times are
both $\SI{2.2\pm 0.1}{\milli\second}$
and agree to within the standard error on the fits.
} 
\label{fig:deviceA2_parity_measurements}
\end{figure*}

\Cref{fig:deviceA2_parity_measurements} presents experimental data from measurement A2, which
is another measurement
of device A in the same cooldown.
Measurement A2 produced
a data set which is similar to A1, indicating the
reproducibility of our data and the device’s stability from
one measurement run to another.

\begin{figure*}
\includegraphics[width=17.9cm]{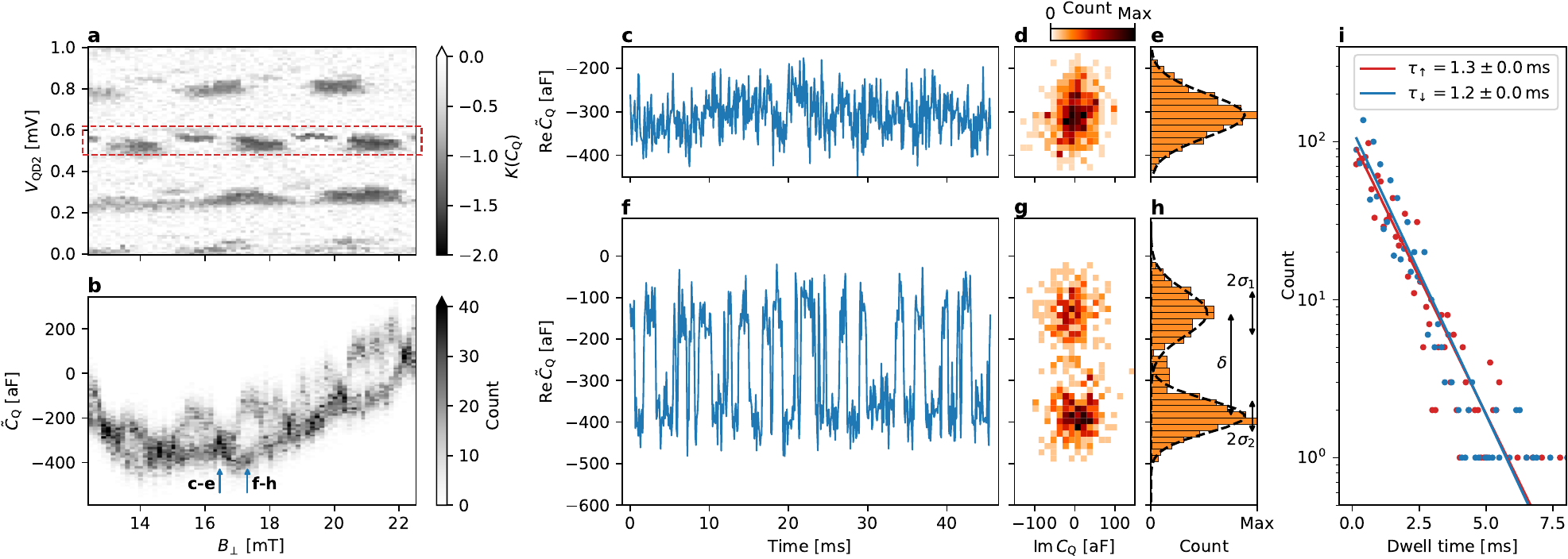}
\vspace{-5mm}
\caption{Measurement B1: 
\figpanel{a}~The measured kurtosis $K(\Cq)$ of dot~2 in device B (measurement B1) as a function of $B_\perp$ and $\VQD{2}$ following the tune-up procedure described in \Cref{sec:device-tune-up}.
\figpanel{b}~A histogram of $\tCq$ values as a function of flux for the $\VQD{2}$ value indicated by the black arrows in panel a, showing bimodality that is flux-dependent with period $h/2e$.
\figpanel{c,f}~Time traces at the two flux values marked by the vertical arrows in panel b, corresponding to minimal (panel c) and maximal (panel f) $\Delta \Cq$.
\figpanel{d,g}~The raw rf signal converted to complex $\tCq$ by the method described in \Cref{sec:cq_conversion} for the time trace shown in panels c and f. 
\figpanel{e,h}~Histograms of $\mathrm{Re}\,\tCq$ with Gaussian fits for an extraction of the $\mathrm{SNR} = 2.63$.
\figpanel{i}~A histogram of dwell times aggregated over all values of $B_\perp$ where the signal shows bimodality. Fitting to an exponential shows that the up and down dwell times are, respectively,
$\SI{1.3\pm 0.1}{\milli\second}$
and $\SI{1.2\pm 0.1}{\milli\second}$.
}
\label{fig:deviceB1_parity_measurements}
\end{figure*}

\Cref{fig:deviceB1_parity_measurements} shows data from device B (measurement B1) made in a different dilution refrigerator with the same wiring configuration. The same tune-up and measurement procedures are used for both devices and the data are qualitatively similar. \Cref{tab:snr_params} compares the parameters extracted from fitting the time traces in device B to those measured with device A. In run B1, the magnitude of the $\Cq$ response is reduced relative to the runs on device A.
Applying the method discussed in \Cref{sec:interferometry_regimes} to analyze the data quantitatively, we find good agreement for $\tm{1}\approx \tm{2} \approx \SI{3}{\ueV}$, $t_{12} \approx t_{23} \approx \SI{16}{\ueV}$, and $\EM \approx \SI{2}{\ueV}$. The reduced quantum capacitance can thus be attributed to reduced values of $\tm{1}$ and $\tm{2}$ compared to what was achieved in measurement A1.

Another difference from the measurements on device A is the reduction in $\sigma_{1,2}$ by approximately a factor of two.  This reduction is consistent with the relative readout amplitude used for the measurements on the two devices, with the amplitude for device B being larger by a factor of two.  Both of these differences indicate that further optimization of the readout is possible, enabling higher SNRs in future experiments.

\begin{table}
\begin{center}
\begin{tabularx}{\columnwidth}{|c|c|c|c|c|c|}
\cline{1-6}
Device and run &
$\Delta\Cq$ [aF] &
$\sigma_1$ [fF] &
$\sigma_2$ [fF] &
SNR &
$\tauRTS$ [ms] \\ 
\cline{1-6}
A1 &
\qty{962}{} &
\qty{101}{} &
\qty{92}{} &
\qty{5.0}{} &
\qty{2.0}{} \\ 
A2 &
\qty{701}{} &
\qty{96}{} &
\qty{104}{} &
\qty{3.5}{} &
\qty{2.2}{} \\
B1 &
\qty{250}{} &
\qty{54}{} &
\qty{39}{} &
\qty{2.63}{} &
\qty{1.3}{} \\
\cline{1-6}
\end{tabularx}
\end{center}
\vskip -3mm
\caption{Fitted parameters from parity measurements on devices A and B.}
\label{tab:snr_params}
\end{table}

\subsection{Measurements with triple-dot isolated from nanowire}
\label{sec:cutloop}

\begin{figure}
\centering
\includegraphics[width=8.5cm]{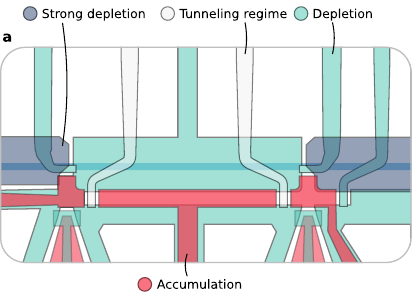}
\includegraphics[width=8.5cm]{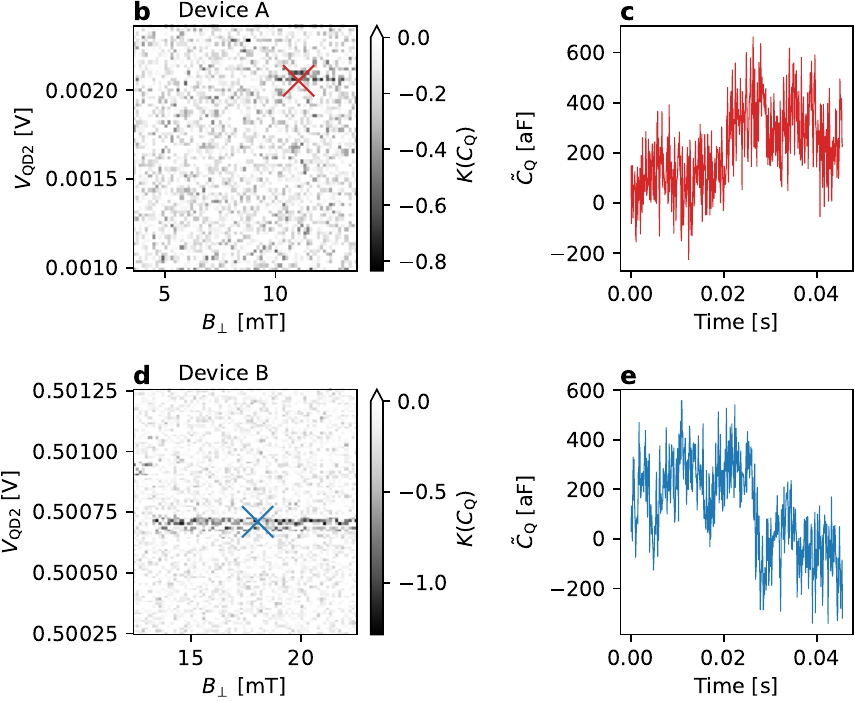}
\vskip -2mm
\caption{
Additional measurements with the two junctions TG1 and TG2 fully depleted.
\figpanel{a}~Gate voltage configuration.
\figpanel{b}~Kurtosis of $\Cq$ time trace $K(\Cq)$ as a function of $B_\perp$ and $\VQD{2}$ with TG1 and TG2 fully depleted in device A. 
In this regime, we observe $K(\Cq) \approx 0$, indicating Gaussian noise, except for the region marked by the red $\times$.  
\figpanel{c}~Time record taken at the position denoted by the red $\times$ in panel b.  
The step in the middle of the dataset accounts for the finite kurtosis seen in panel b but is clearly different from the RTS observed in interferometry measurements.
\figpanel{d,e}~Corresponding data taken from device B, where bimodality is completely absent in this regime.
}
\label{fig:cut_loop}
\end{figure}

To substantiate that the observed RTS is due to the interferometer loop formed between the triple quantum dot and the wire, we perform a similar measurement but with the two gates controlling the coupling between dot~1 and dot~3 and the wire (TG1 and TG2) set to completely deplete the respective junction area as illustrated in \Cref{fig:cut_loop}a. The results of this measurement are shown in \Cref{fig:cut_loop}b-e for a representative value of WP1 for devices A and B. We observe that the kurtosis of the time traces exhibits very little structure and is near zero throughout the measurement, indicating absence of an RTS.  Regions with finite kurtosis (\Cref{fig:cut_loop}c,e) show single steps in the measured $\tCq$, presumably due to low frequency charge noise.    

\subsection{Low field regime}
\label{sec:trivial_measurement}

\begin{figure}
\includegraphics[width=8.7cm]{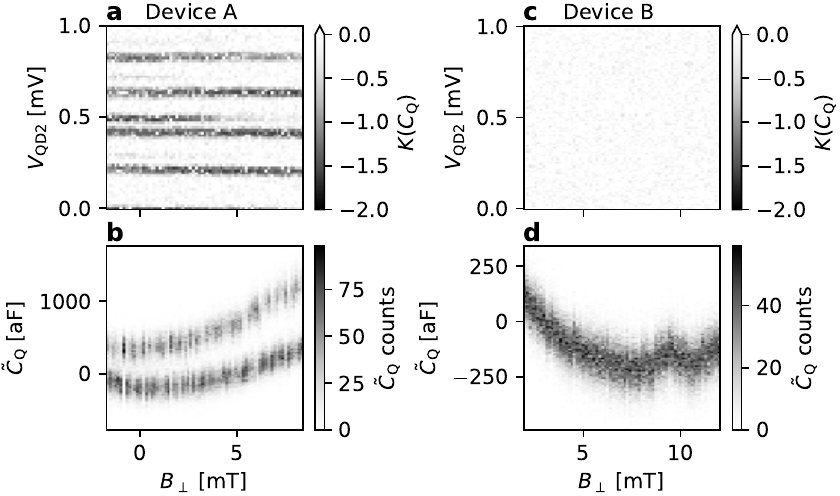}
\vskip -2mm
\caption{
Interferometry measurement at low field.
\figpanel{a}~Kurtosis of $\Cq$ time trace as a function of $B_\perp$ and $\VQD{2}$ at an in-plane field of \SI{0.8}{T} in device A.  
\figpanel{b}~Histogram of $\tCq$ as a function of $B_\perp$, illustrating a flux-independent bimodality. 
\figpanel{c,d}~Corresponding data from device B.
}
\label{fig:trivial_measurement}
\end{figure}

We now address the question of what is observed outside the $\Bpara$ and $\VWP$ regime where a topological phase can be expected. 
We perform interferometry measurements at low magnetic field ($\SI{0.8}{T}$), well before the gap in the nanowire has closed, to investigate the signal that is observed in generic gapped regimes.
As shown in \Cref{fig:trivial_measurement}, we observe an absence of flux-$h/e$ periodicity, as expected for a gapped wire.

When the gapped state has local sub-gap states at the junctions, the $\tCq$ time record displays bimodality.
We interpret this bimodality as arising from the quantum dots coupling to a local subgap state which is poisoned by quasiparticles in the nanowire.
When the gapped state does not have local sub-gap states at the junctions, the bimodality is gone, too.
In short, bimodality can be observed whenever there are subgap states that are poisoned, but flux-$h/e$ periodicity requires a single state to be coupled to both small dots in order to close the interference loop.

\begin{figure}
\includegraphics[width=8.0cm]{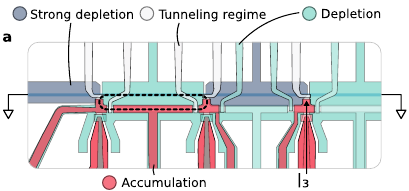}
\includegraphics[width=8.5cm]{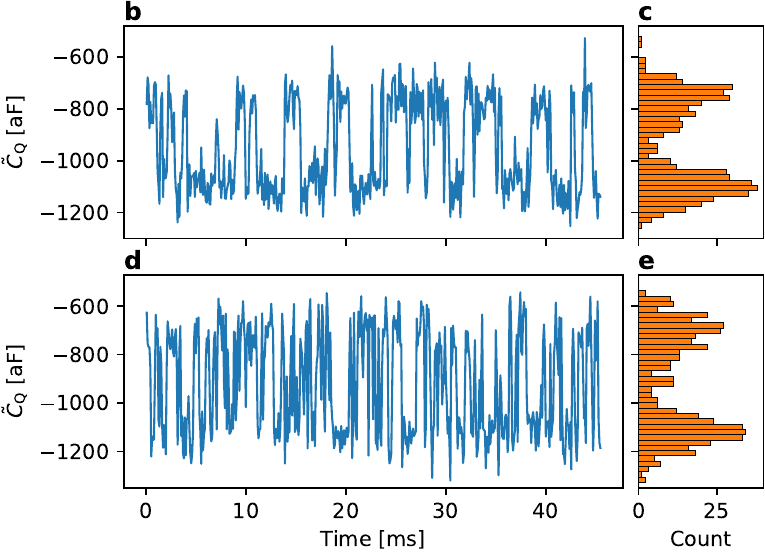}
\vspace{-1mm}
\caption{
\figpanel{a}~Tuning configuration for quasiparticle injection measurement in device B. 
The left interferometer loop is in the standard measurement configuration but now the gates around junction 3 are tuned to form a tunnel junction to enable injection of current $I_3$ at an energy set by bias $V_3$.
$\tCq$ time series and histograms for $\SI{65}{\micro\volt}$ (b,c) and $\SI{85}{\micro\volt}$ (d,e) injector bias, illustrating bimodality with dwell times modulated by the injector bias.
}
\label{fig:deviceB_telegraph}
\end{figure}

\subsection{Quasiparticle injection}
\label{sec:qp_injection}

In \Cref{fig:deviceB_telegraph}a, we show
the device configuration used to inject
quasiparticles into the interference loop.
These gate settings tune the device so that
it realizes the schematic shown in
\Cref{fig:deviceB_dwell_times}a.
In \Cref{fig:deviceB_telegraph}, we show
$\Delta\Cq$ time traces for bias voltages
$\SI{65}{\micro\volt}$ (panels b,c) and $\SI{85}{\micro\volt}$ (panels d,e).
These time traces yield two of the
data points on the plot shown in
\Cref{fig:deviceB_dwell_times}b.


\section{Charge noise}
\label{sec:charge_noise}

\begin{figure}
\includegraphics[width=\columnwidth]{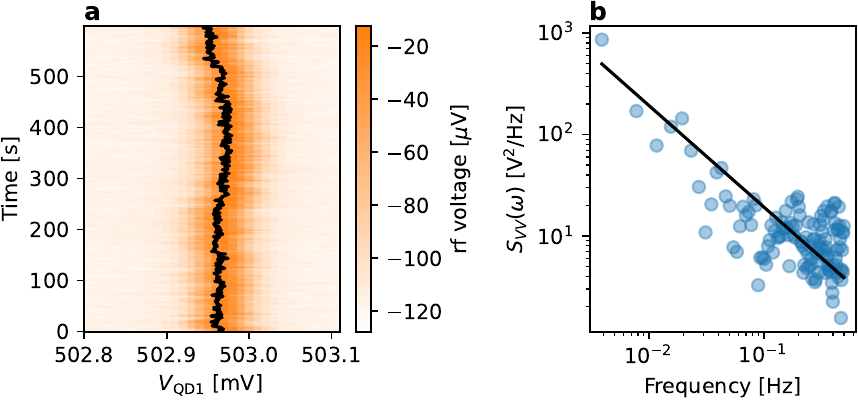}
\vskip -2mm
\caption{
Measurements of low-frequency charge noise in an isolated triple-dot configuration.
\figpanel{a}~Dispersive gate sensing of the linear dot QD2 as a side dot QD1 is repeatedly swept through resonance with the linear dot. The other side quantum dot QD2 is detuned for this measurement.  Tracking the peak of the measured rf voltage (black line) produces a time-trace of the voltage at resonance. Plotted data is from device B. 
\figpanel{b}~Power spectrum of the voltage at resonance, computed from the time trace in panel a.
Line is a fit to a $1/f$ model $\Svv(\omega) = V_0^2/\omega$.
}
\label{fig:charge_noise}
\end{figure}

To measure the low-frequency charge noise in our devices we use the isolated triple-dot system described in~\Cref{sec:dot-tuning}. This is done by detuning one of the side dots from resonance using the charge stability diagram in~\Cref{fig:dot_tuneup}d.
The plunger gate of the other side dot is then swept through resonance with the linear dot while dispersively sensing the gate response of the linear dot.  Repeating this measurement for approximately 10 minutes, we extract the plunger voltage at which the resonance appears as a function of time (black line in
\Cref{fig:charge_noise}a).
We Fourier transform this time trace to compute the power spectral density $\Svv(\omega)$ of the fluctuations in the plunger voltage at which the resonance peak appears (\Cref{fig:charge_noise}b).
Finally, we fit a $1/f$ noise model $\Svv(\omega) = V_0^2/\omega$ to the power spectrum to extract $V_0$.

We relate $V_0$ to the charge noise on the dots by making the assumption that the fluctuations in the chemical potentials of the side and linear dots are uncorrelated. The voltage spectral density is then
\begin{eqnarray}
    \Svv(\omega) = \frac{S_\mathrm{s}(\omega)+ S_\mathrm{l}(\omega)}{e^2 \alpha_\mathrm{s}^2},
\end{eqnarray}
where $e$ is the electron charge, $S(\omega)$ characterizes fluctuations in the chemical potentials of the dots (subscripts denote the side and linear dot), and $\alpha$ is the lever arm.  A scaling argument may then be used to connect the spectral densities of the dots with their respective areas $A$ and through those their charging energies~\cite{Paladino14}:
\begin{eqnarray}
    \frac{S_\mathrm{s}(\omega)}{S_\mathrm{l}(\omega)} = \frac{A_\mathrm{l}}{A_\mathrm{s}} = \sqrt{\frac{\EC^\mathrm{s}}{\EC^\mathrm{l}}}. 
\end{eqnarray}
The charge noise on the linear dot may then be written in terms of $\Svv(\omega)$
\begin{eqnarray}
    S_\mathrm{l}(\omega) = \frac{e^2\alpha_\mathrm{s}^2}{1+\sqrt{\EC^\mathrm{s}/\EC^\mathrm{l}}}
    \Svv(\omega).
    \label{eq:lin_dot_charge_noise}
\end{eqnarray}

Modelling the charge noise on the linear dot with a $1/f$ spectrum $S_\mathrm{l}(\omega) = S_0/\omega$, we compute $S_0$ for measurements made with the left side quantum dot QD1 and the right side quantum dot QD3 on devices A and B (\Cref{tab:charge_noise}).
We extract $\sqrt{S_0}$ between 1 and \qty{2}{\ueV} across both devices and dots.

\begin{table}
\centering
\begin{tabular}{ |l|c|c|} 
 \hline
  Quantum dot & $\EC$ [\qty{}{\ueV}]& $\alpha$  \\
  \hline
 QD1 & 140 & 0.46\\ 
 QD2 & 45 & 0.45\\ 
 QD3 & 100 & 0.48\\ 
 \hline
\end{tabular}
\caption{
Measured parameters for the quantum dots in the isolated triple-dot configuration.  
As discussed in \Cref{sec:toy-model-sims}, the $\EC$ values are renormalized from the single dot charging energies quoted in \Cref{fig:dot_tuneup}.
}
\label{tab:charge_noise_inputs}
\end{table}

\begin{table}
\centering
\begin{tabular}{ |l|c|c|} 
 \hline
  Device & $\sqrt{S_0}$ from QD1 [\qty{}{\ueV}]& $\sqrt{S_0}$ from QD3 [\qty{}{\ueV}]\\
  \hline
 A & \qty{1.8\pm0.1}{} &  \qty{1.2\pm0.1}{} \\ 
 B & \qty{0.96\pm0.03}{} & \qty{1.39\pm0.05}{}\\ 
 \hline
\end{tabular}
\caption{
Results of charge noise measurements on the linear quantum dot for three devices. $S_0$ is computed with \Cref{eq:lin_dot_charge_noise} using the parameters in~\Cref{tab:charge_noise_inputs} and the fitted values of $V_0$.  
The two columns show measurements inferred with the left and right side dots QD1 and QD3.
}
\label{tab:charge_noise}
\end{table}

\section{Electron temperature}
\label{sec:electron_temperature}

\begin{figure}
\includegraphics[width=\columnwidth]{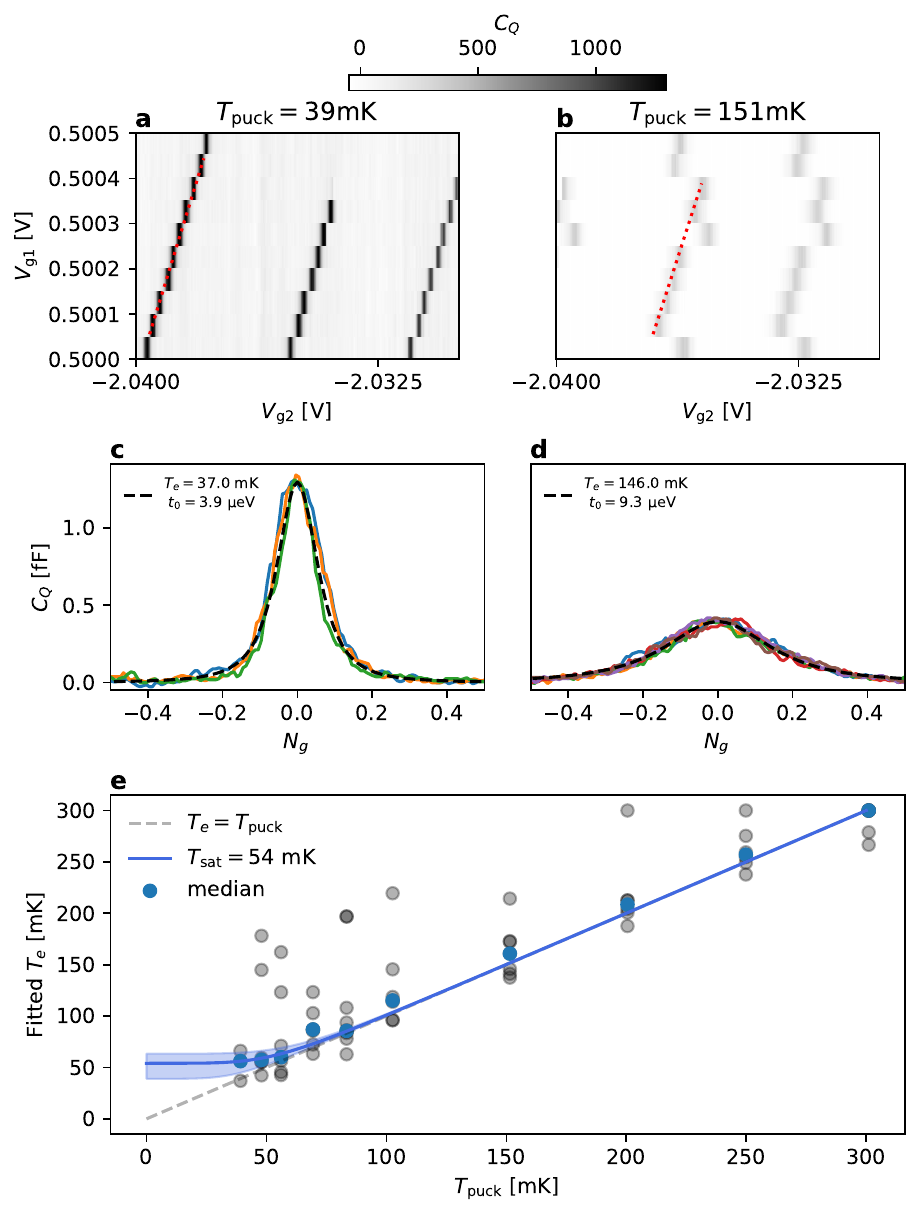}
\vspace{-6mm}
\caption{
Thermometry on device B.
\figpanel{a,b}~Gate-gate maps taken at two different puck temperatures. 
Each line corresponds to a different charge transition of the double-dot system. 
For simplicity of the figure, we focus on the charge transition highlighted with the dotted red line.
\figpanel{c}~Each solid colored line here represents a horizontal cut across the dotted red line in panel a. 
The black dashed line represents a fit to simulation results with the parameters indicated in the legend.
\figpanel{d}~Same as panel c for the transition line highlighted in panel b.
\figpanel{e}~Each grey circle shows the extracted temperature for all the transition lines in one of the gate-gate maps. 
Different grey circles for a fixed puck temperature correspond to different choice of the junction gate voltage and therefore different coupling $t_0$. 
The blue line shows a fit of \Cref{eqn:therm_scaling} to the median extracted value for each puck temperature (yielding $T_\mathrm{sat} = \SI{54}{\milli\kelvin}$) while the shaded area is given by a fit to the upper and lower quartile with corresponding $T_\mathrm{sat} = [\SI{39}{\milli\kelvin},\SI{63}{\milli\kelvin}]$. 
We excluded outliers that differ from $T_\mathrm{puck}$ by more than \SI{100}{\milli\kelvin} from the fit.
}
\label{fig:deviceB_thermometry}
\end{figure}

Since the temperature is a key input parameter to our simulations, we extract the electron temperature $T_\mathrm{e}$ from fitting the $\Cq$ response of a double quantum dot (DQD) formed in our parity measurement devices to simulations. 
By tuning to the simple DQD configuration, we can reduce the number of fit parameters and thus obtain a reliable estimate of the electron temperature in a setup where all leads are disconnected, as they are in the parity readout. 
Experimentally, we minimize the coupling between the TQD and the superconducting wire by setting TG1 and TG2 far below their threshold voltage. 
Furthermore, we deplete one of the smaller dots (dot~1 or dot~3) by setting its plunger value slightly below depletion. 
For sake of concreteness, we will focus on the scenario where dot~3 is depleted, and we are thus left with a DQD configuration formed by dots 1 and 2. 
This is the configuration used for the measurement on device B shown in \Cref{fig:deviceB_thermometry}.

In this configuration, we can tune the coupling between the two quantum dots using the voltage applied on QC1. 
We then measure $\Cq$ on the long quantum dot as function of the plunger on each of the two quantum dots, thus yielding gate-gate maps as shown in \Cref{fig:deviceB_thermometry}a. 
We process a given gate-gate map by algorithmically identifying lines of $\Cq$ peaks and fitting each $\Cq$ peak to a simulation of a simple DQD system using the simulation framework described in \Cref{sec:dynamical-cq}, as shown for some example cuts in \Cref{fig:deviceB_thermometry}c,d. 
This is repeated for different values of QC1 to cover a range of couplings between the quantum dots. 
We restrict the fit by assuming that (i)~all $\Cq$ peaks for fixed fridge temperature and QD cutter setting can be fit using the same electron temperature $T_\mathrm{e}$, (ii)~all $\Cq$ peaks along a given line share the same coupling between the two quantum dots and to the bath. 
However, we allow these parameters to vary between lines to account for variation in the matrix elements between quantum dot levels. 
Finally, for each line trace, we allow a shift in $\Ng{i}$ to account for low-frequency charge noise and finite $\Ng{i}$ resolution of the measured data.

A key parameter for this approach is the lever arm. 
While we can extract an initial value from Coulomb diamonds recorded during tune-up of the device (cf \Cref{sec:dot-tuning}), we can more accurately calibrate the relevant lever arm in the DQD configuration by varying the fridge temperature using the mixing-chamber heater and adjusting the lever arm such that at high temperatures, the extracted temperature matches the fridge temperature. The result of this is shown in \Cref{fig:deviceB_thermometry}e. 
We fit the extracted temperatures to
\begin{equation} \label{eqn:therm_scaling}
T_\mathrm{e} = (T_\mathrm{puck}^p + T_\mathrm{sat}^p)^{1/p},
\end{equation}
where $T_\mathrm{e}$ is the extracted electron temperature at a given fridge (in this case, puck) temperature $T_\mathrm{puck}$, and $T_\mathrm{sat}$ is the temperature to which the electrons saturate at the lowest fridge temperatures~\cite{Meschke02, Casparis2015}. 
Here, we fix $p = 5$, as appropriate for electron-phonon cooling being the dominant physical process. 
From this, we obtain an estimate of the electron temperature of $\SIrange{40}{60}{\milli\kelvin}$ at base temperature of the fridge.

\section{Measurement of non-equilibrium quasiparticle density in a Cooper pair box device}
\label{sec:QPP}

\begin{figure*}
{\includegraphics[width=5.6cm]{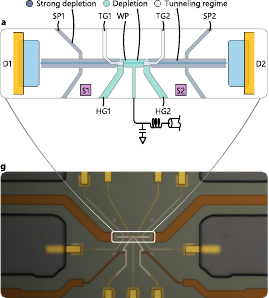}\includegraphics[width=8.0cm]{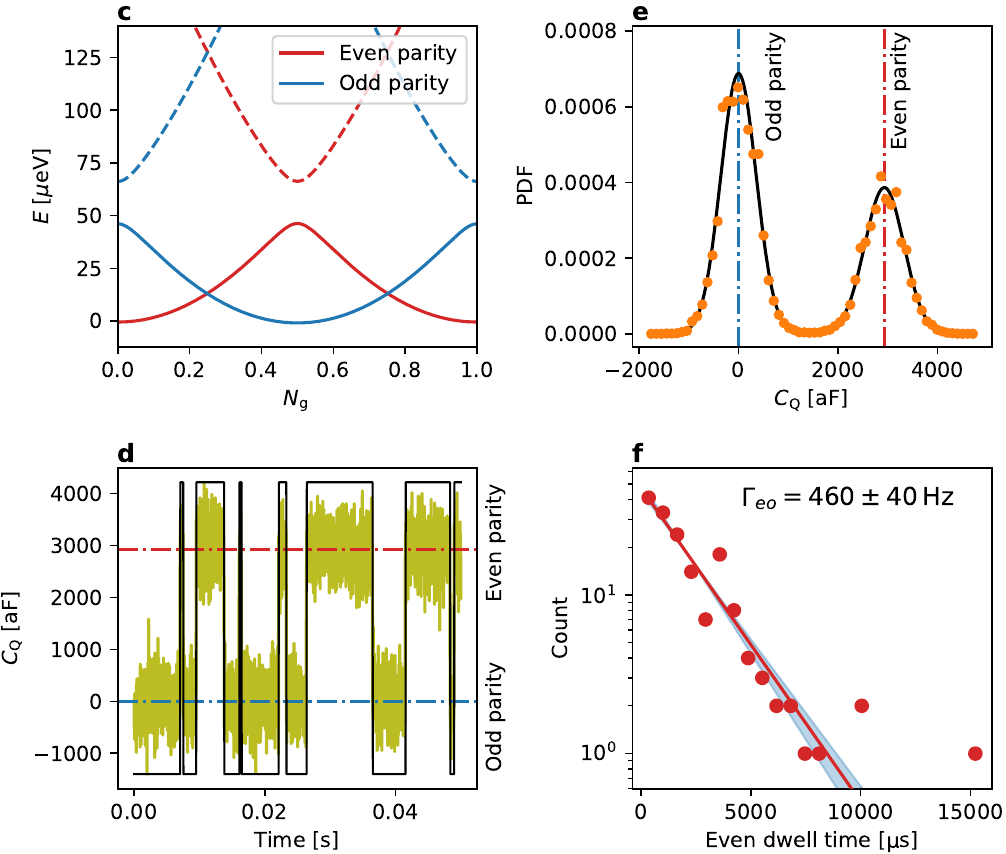}\includegraphics[width=3.9cm]{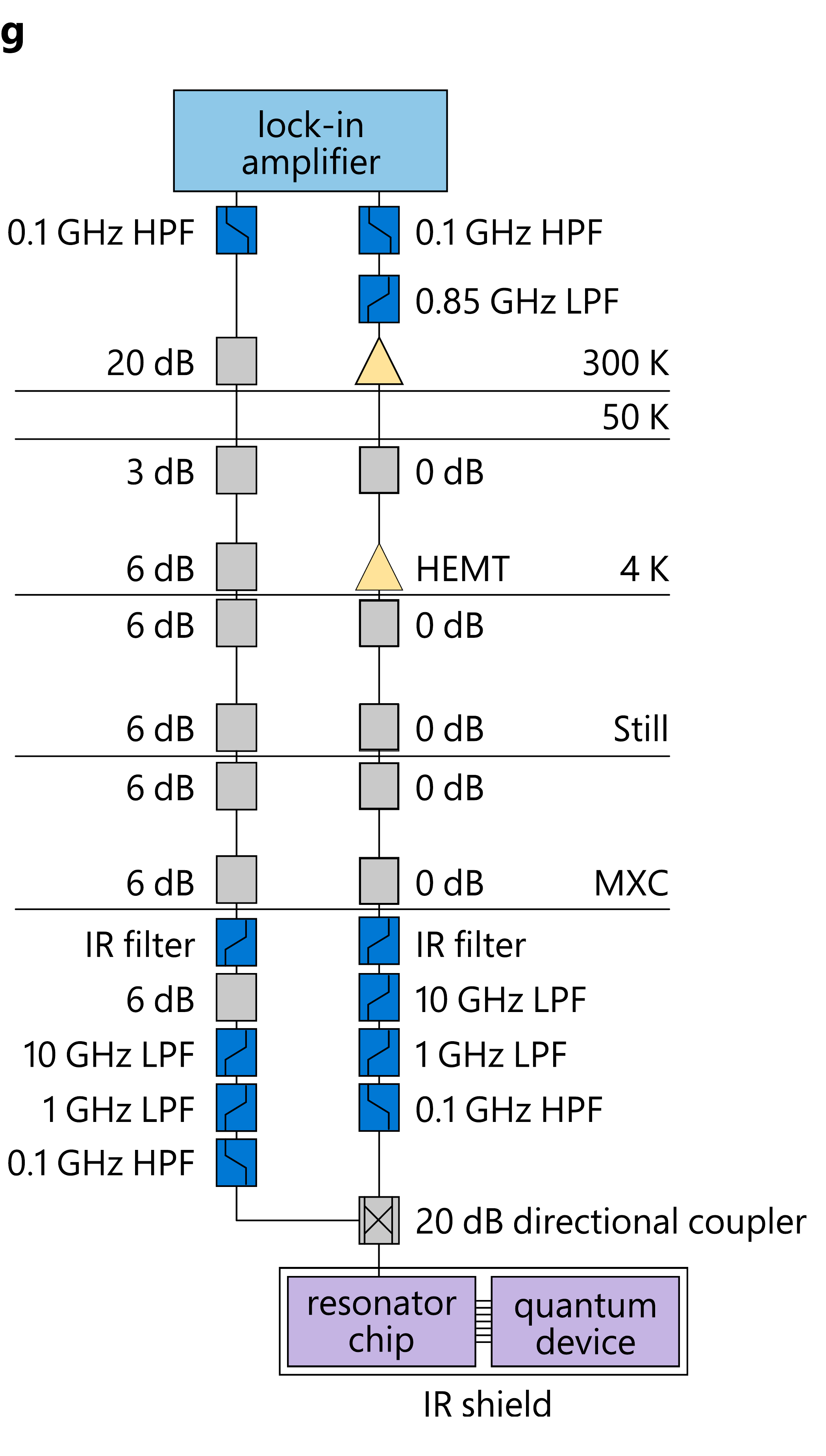}}
\caption{
Cooper pair box (CPB) quasiparticle poisoning measurement at $B=0$.
\figpanel{a}~Schematic of the CPB device, with the resonator connected to the CPB plunger gate.
\figpanel{b}~An optical image of an equivalent device. 
\figpanel{c}~Energy levels of the island, with even(odd) parity branches marked in red(blue), and excited states (dashed lines). 
\figpanel{d}~Section of a representative time trace showing the extracted quantum capacitance shift. Quasiparticle poisoning events are clearly visible, with the extracted parity marked in black. 
\figpanel{e}~The probability distribution function (PDF) of $\Cq$ values across a time trace. The SNR of this measurement is 3.8. 
\figpanel{f}~Distribution of even dwell times and an exponential fit to the data, showing a poisoning rate of \qty{460\pm 40}{\Hz}. 1$\sigma$ confidence interval is marked in blue.
\figpanel{g}~Schematic of the readout chain connected to the resonator.
}
\label{fig:qpp}
\end{figure*}

To gain a deeper understanding of the quasiparticle poisoning phenomenon in our devices and to extract the non-equilibrium quasiparticle density we have designed and fabricated an auxiliary device. 
This device consists of a Cooper pair box (CPB), which is coupled to superconducting leads via two gate-controlled Josephson junctions, as illustrated in \Cref{fig:qpp}a. 
The device is fabricated on a comparable InAs/InAlAs heterostructure, maintaining the same wire width to ensure that the cross-section of the device matches that of the TQDI device. 
The plunger gate for the CPB is coupled to a resonator similar to that of the TQDI device, the parameters of which are listed in \Cref{tab:qpp}. 
This setup allows for the direct measurement of the island’s parity by observing the dispersive shift in the time domain, as outlined in \Cref{sec:readout_system} and discussed in Ref.~\onlinecite{Shaw08}.  

\begin{table}
\centering
\begin{tabular}{ |l|c|c| } 
 \hline
 Param. & Description & Value \\ 
 \hline\hline
 $f_0$ & Resonator frequency & \qty{776.5}{\mega\hertz} \\ 
 $\kappa_\mathrm{ext}/2\pi$ & Resonator readout coupling & \qty{14.8}{\mega\hertz} \\
 $\kappa_\mathrm{int}/2\pi$ & Resonator internal loss & $<\SI{1.5}{\mega\hertz}$ \\ 
 \hline
 $\Delta_0$ & Parent superconducting gap & \qty{275}{\ueV} \\
 $\EC$ & Charging energy & \qty{225}{\ueV} \\
 $\EJ$ & Josephson energy & \qty{20}{\ueV} \\
 $\alpha$ & Gate lever arm & 0.8 \\
 $g$ & NS dimensionless conductance & 0.3 \\
 $\nu_\mathrm{Al}$ & Normal DOS of aluminum & \qty{3e4}{\per\ueV\per\micro\meter\cubed} \\
 \hline
\end{tabular}
\caption{Measured device parameters for the Cooper pair box device.}
\label{tab:qpp}
\end{table}

During the measurement, the device is operated in the Cooper pair box regime where one of the Josephson junctions is closed. The measured parameters of the device are summarized in \Cref{tab:qpp}. 
The device is then tuned to an odd offset charge which we refer to as $\Ng{i} = 1$ which leads to a degeneracy of the $N = 0$ and $N = 2$ state of the Cooper pair box. 
The finite Josephson energy then leads to an avoided crossing which gives rise to a quantum capacitance
\begin{equation}
\Cq^\CPB = \frac{
    8e^2 \alpha^2 \EJ^2 \tanh\Bigl(
        \sqrt{(2\alpha eV_g)^2 + 4 \EJ^2} / 2 \kB T
    \Bigr)
}{
    \bigl[(2 \alpha e V_g)^2 + 4 \EJ^2\bigr]^{3/2}
}.
\label{eq:CQ_CPB}
\end{equation}
Here $V_g$ and $\alpha$ are the plunger gate voltage and corresponding lever arm $\alpha$ of the Cooper pair box, respectively. 
When the device is in the odd state due to a quasiparticle poisoning event, the charge states differing by a Cooper pair are no longer resonant and the quantum capacitance vanishes. 
The parity-dependence of the quantum capacitance then leads to a random telegraph signal, from which, by continuously observing the capacitive response, we can directly extract the poisoning rate. 
The sample enclosure and dc filtering are chosen to be equivalent to the TQDI device, however the readout circuit utilizes a Caltech CITLF3 amplifier and places the IR filters before the coupler, at the mixing chamber stage. 
A schematic of the circuit is shown in \Cref{fig:qpp}g. 
We expect that the differences between the readout circuits are negligible, or will degrade performance relative to the configuration in the main experiment, meaning the extracted number here will act as an upper bound on the quasiparticle density. 
The measurement and subsequent extraction of parity closely follows the procedure detailed in Ref.~\onlinecite{Shaw08}, with a filter bandwidth of \qty{100}{\kHz}, much shorter than the extracted time constant of quasiparticle poisoning. 
A subsection of a representative RTS trace, with the readout signal converted to units of $\Cq$ following the method detailed in \Cref{sec:cq_conversion}, is shown in \Cref{fig:qpp}d. 
We assign the even/odd parity states based on thresholding the $\Cq$ signal. 
Whenever the signal exceeds the mean of the even parity we label it as even and keep that label until it drops below the mean of odd parity state. 
This method is also used to label the data in \Cref{fig:deviceA1_parity_measurements}e  and works well in suppressing state mislabeling due to finite SNR.%
\footnote{In addition, for the labeling of the data used in \Cref{fig:deviceA1_parity_measurements}e, one needs to take into account that there the assignment of high/low values of $\Cq$ for the even/odd parities switches as a function of flux with an $h/e$ periodicity.} 
Finally, we extract $\GammaEO = \qty{460\pm40}{\Hz}$ by fitting the even-state dwell-time distribution to an exponential distribution as shown in \Cref{fig:qpp}f. 

The extracted even-to-odd switching rate $\GammaEO$ can be connected~\cite{Shaw08} to the quasiparticle density via 
\begin{equation}
\GammaEO 
= \frac{g}{4 \pi}
\frac{n_\mathrm{qp}^\CPB}{\nu_\mathrm{Al}}
\sqrt{\frac{\delta E}{2\Delta_0}},
\label{eq:qp_density}
\end{equation}
where $g$ is the normal state dimensionless conductance of the Josephson junction, $\delta E = \EC - \EJ/2$, $\Delta_0$ is the (parent) superconducting gap, and $\nu_\mathrm{Al}$ is the Al density of states at the Fermi level. 
We extract the charging energy $\EC$ and lever arm $\alpha$ from Coulomb diamonds (not shown). 
We can then use \Cref{eq:CQ_CPB} to fit the width of the peak in $\Cq^\CPB$ when changing the gate voltage of the Cooper pair box to determine $\EJ$, which yields $g = 4\EJ/\Delta_0$. 
With these parameters we find using \Cref{eq:qp_density} $n_\mathrm{qp}^\CPB \approx \qty{0.6}{\per\micro\meter\cubed}$ at $B = 0$, within an order of magnitude of the inferred value in the TQDI device of $n_\mathrm{qp} \approx \qty{1}{\per\micro\meter\cubed}$ at $B = \SI{2}{\tesla}$.

\end{document}